\documentclass[preprint,aps,showpacs,superscriptaddress,11pt]{revtex4-1}
\usepackage[utf8]{inputenc}
\usepackage[a4paper, portrait, margin=1in]{geometry}
\usepackage{graphicx}
\usepackage{natbib}
\usepackage{amsmath,amssymb}
\usepackage[title]{appendix}

\usepackage{setspace}

\newcommand{\hbAppendixPrefix}{A}
\newcommand{\order}{s}

\newcommand{\noise}{\lambda_c(\epsilon)}

\begin{document}
\title{\textbf{Dynamical Ising model of spatially-coupled ecological oscillators}}


\author{Vahini Reddy Nareddy}
\email{vnareddy@umass.edu}
\affiliation{Department of Physics, University of Massachusetts,
Amherst, Massachusetts 01003 USA}
\author{Jonathan Machta}
\email{machta@physics.umass.edu}
\affiliation{Department of Physics, University of Massachusetts,
Amherst, Massachusetts 01003 USA}
\affiliation{Santa Fe Institute, 1399 Hyde Park Road, Santa Fe, New Mexico
87501, USA}
\author{Karen C.~Abbott}
\email{kcabbott@case.edu}
\affiliation{Department of Biology, 10900 Euclid Ave, Cleveland, OH 44106, USA}
\author{Shadisadat Esmaeili}
\email{sesmaeili@ucdavis.edu}
\affiliation{Department of Environmental Science and Policy, One Shields Avenue, University of California, Davis, CA 95616, USA}
\author{Alan Hastings}
\email{amhastings@ucdavis.edu}
\affiliation{Department of Environmental Science and Policy, One Shields Avenue, University of California, Davis, CA 95616, USA}
\affiliation{Santa Fe Institute, 1399 Hyde Park Road, Santa Fe, New Mexico
87501, USA}

\begin{abstract}
   Long-range synchrony from short-range interactions is a familiar pattern in  biological and physical systems, many of which share a common set of ``universal'' properties at the point of synchronization.  Common biological systems of coupled oscillators have been shown to be members of the Ising universality class, meaning that the very simple Ising model replicates certain spatial statistics of these systems at stationarity.  This observation is useful because it reveals which aspects of spatial pattern arise independently of the details governing local dynamics, resulting in both deeper understanding of and a simpler baseline model for biological synchrony. However, in many situations a system's dynamics are of greater interest than their static spatial properties.  Here, we ask whether a dynamical Ising model can replicate universal and non-universal features of ecological systems, using noisy coupled metapopulation models with two-cycle dynamics as a case study.  The standard Ising model makes unrealistic dynamical predictions, but the Ising model with memory corrects this by using an additional parameter to reflect the tendency for local dynamics to maintain their phase of oscillation. By fitting the two parameters of the Ising model with memory to simulated ecological dynamics, we assess the correspondence between the Ising and ecological models in several of their features (location of the critical boundary in parameter space between synchronous and asynchronous dynamics, probability of local phase changes, and ability to predict future dynamics).  We find that the Ising model with memory is reasonably good at representing these properties of ecological metapopulations.  The correspondence between these models creates the potential for the simple and well-known Ising class of models to become a valuable tool for understanding complex biological systems. 
  
\end{abstract}
\maketitle

\setstretch{1.0}
\section{Introduction}

Synchrony of dynamics in spatially extended systems
has been a subject of intense study in a diverse array of scientific disciplines and range of biological scales \citep{Kuramoto:1975ux, Acebron:2005td, Bjornstad:1999un, Strogatz:2003we, Liebhold:2004ba, Harris2015}. 
In  ecological systems, the study of synchrony of oscillations in population numbers across space and time has a long history \citep{Moran:1953tr} and has provided great insights into fundamental issues of population dynamics~\citep{Bjornstad:1999un, Liebhold:2004ba}.  
The dynamics of flocking in birds or schooling in fish and similar behavior in bacteria~\citep{Zhang13626} essentially are examples of synchrony across a large scale determined by local interactions~\citep{sumpter2008information, Toner1995, Bialek:2012wq}.  At the suborganism scale, how the dynamics of individual neurons lead to collective behavior is a key question in neuroscience \citep{10.3389/fncom.2015.00069}.  The synchrony of neural oscillators is thought to play an important role in behavior \citep{Harris2015} and in various pathologies \citep{uhlhaas2006neural}.

Given the ubiquity of analogous synchronous behavior across a range of biological systems and scales, it is reasonable to look for explanations that do not depend on fine details.  One approach is based on the Ising model from physics which is an idealized description of the macroscopic behavior of magnetic materials based on the local coupling of the microscopic magnetic moments of electrons.  The Ising model (see, for example, \citep{Stanley71, GoldenfeldNigel1992Lopt, Cardy:1996th, Sethna:2006wn, Sole:2011us}) is  a fundamental model in statistical physics and the simplest example of a ``spin'' model. It was introduced to understand how long-range order can develop from local interactions in the setting of ferromagnetism -- the global alignment of atomic-scale magnetic moments to produce bulk magnetism in materials such as iron.  A variety of biological synchrony phenomena have previously been described using models based on the Ising model and related spin models, such as 
cell synchrony  \citep{Weber2016},  pattern formation  \citep{Wang1980},  synchronous nerve firing \citep{Cocco14058, Schneidman:2006he},  swarming and flocking dynamics   \citep{calovi2014swarming, Bialek:2012wq}, and  masting behavior \citep{Noble:2018ge}.  These previous studies have focused on static properties, but  
synchronization is a dynamic phenomenon. Important questions emerge when considering the dynamics of synchrony  that require further properties of the Ising model.  We focus here on ecological systems where these dynamic features emerge, but the ideas and approaches should have much wider applicability.

Long-range synchrony in population fluctuations  is widely observed and is thought to occur through several non-mutually-exclusive mechanisms \citep{Grenfell:1998vv, Bjornstad:1999un, Liebhold:2004ba, Vasseur:2009dx, Abbott:2011fv}.  Long-range correlations in environmental perturbations and long distance dispersal are both capable of synchronizing local population dynamics across a large geographic scale \citep{Moran:1953tr, Ripa:2000vk, Greenman:2001vf, Desharnais:2018er, Hopson:2018im}.  However, some species with limited dispersal abilities also show strong spatial synchrony \citep{Peltonen:2002wb}.  For populations with cyclic dynamics, short-range dispersal (or, equivalently, other forms of local coupling such as resource sharing between neighboring trees \citep{Noble:2018ge}) is often sufficient to drive long distance synchrony \citep{Jansen:1999va, Bjornstad:2000vb, Earn:2000fm, Fox:2011dk, Haynes:2019cw}, suggesting a role  for short distance coupling in  some of the examples of synchrony we see in nature.

 Population densities of species such as forest insects \citep{Turchin:1990wh}, voles \citep{Barraquand:2014bx}, and annual plants \citep{Symonides:1986wl}, as well as fruit yield of alternate-bearing plants \citep{Noble2015}, may exhibit a strong pattern of alternation between high and low states.  This prevailing 2-cycle has two possible phases of oscillations:  highs in odd or highs in even years.  This suggests a correspondence  to the discrete ``spin up'' or ``spin down'' states of  the Ising model.  Environmental stochasticity can explain the fact that exact 
  highs and lows vary from cycle to cycle in real populations, and that real 2-cycles may at times change their phase, corresponding to ``spin flips'' in dynamical Ising models. 

Local coupling of multiple subpopulations undergoing noisy 2-cycles in a metapopulation can lead to long-range synchrony \citep{Bjornstad:2000vb, Abbott:2007eo, Wysham:2007bb, Abbott:2011fv, Noble2015}.  While many continuous measures of synchrony have been proposed \citep{Blasius:1999tp, Bjornstad:2001vb, Buonaccorsi:2001to, Gouhier:2014ff, Walter:2017dl}, we can take advantage of the binary nature of the 2-cycle to classify metapopulations as either ``synchronous'' at the scale of the metapopulation (most subpopulations are high in the same years), or else ``incoherent'' (with perhaps some locally synchronized populations but no long-range synchrony).  Intuitively, higher dispersal rates and weaker (spatially uncorrelated) environmental stochasticity promote synchrony.  

For metapopulations with noisy local 2-cycles, the transition from incoherence to global synchrony occurs abruptly with gradual increases in dispersal or decreases in noise \citep{Noble2015}.  This kind of sharp phase transition is also a characteristic of the Ising model \citep{Sethna:2006wn, GoldenfeldNigel1992Lopt}.
In its original application to magnetic materials, the strength of microscopic 
interaction between electron magnetic moments (``spins'') and the temperature (analogous to noise level) of the system determine whether  spins are globally aligned so that the material displays permanent magnetism. Systems that follow the same power-law scaling of correlation functions as the Ising model at the transition point from disordered to ordered are in, what is known as, the Ising universality class.  In previous work, Noble et al.~showed that many ecological 2-cycle oscillators fall into this class \citep{Noble2015}.

The transition from incoherence to synchrony in any model in the Ising universality class shares some non-trivial features {\it exactly} with the transition from disorder to order in the Ising model.   Universality thus allows us to understand many features of synchrony in models like the noisy coupled ecological 2-cycles by instead studying the Ising model, which is very simple, tractable, well-understood, and amenable to detailed, quantitative mathematical and computational analysis. However, universal properties that are exactly shared by all members of the Ising universality class are limited to the large distance and long time properties of correlation functions measured in the stationary (equilibrium) state near the critical point.

Here we posit that a {\it dynamical} Ising model can accurately represent a much broader array of non-universal features of the behavior exhibited by metapopulations with noisy 2-cycles, including the approach to the stationary state and local dynamical properties.
To obtain a faithful representation, we need to go beyond the simplest dynamical Ising model and add a self-interaction (local memory) term to the dynamical Ising model \citep{Pre}.  This memory term reflects that subpopulations are 
strongly influenced by their own current state, due to local density dependence.
We use inference methods \citep{Decelle:2016} to find the Ising model parameters  that best represents  simulated ecological metapopulation dynamics.

We find surprisingly good agreement between the full metapopulation model and its Ising representation.  
This allows us to use the Ising model to develop new  quantitative predictions and qualitative insights.  For example, the noisy, locally coupled Ricker metapopulation  three parameters: dispersal, noise, and local intrinsic population growth rate.  The interplay between these parameters in determining the dynamics and patterns of synchronization can be rather complex \citep{Lande:1999vf, Bjornstad:2000vb, Liebhold:2006dv}.  The Ising model with memory is a significantly simpler model with only two parameters: an effective local coupling and an effective memory term.  Each of these parameters plays an intuitively clear role in determining both the dynamics and patterns of synchronization.  More importantly, for natural systems for which we lack a validated mechanistic model, the dynamical Ising model provides a simple representation that makes very few assumptions about underlying mechanisms.  Using Ising inference methods, we are able to obtain a reasonable description that has both predictive value and yields qualitative insights.

Our study is primarily focused on dynamics near the critical transition to synchrony, where it is most difficult to model the behavior of a system due to the emergence of multiple length scales and time scales \citep{Sethna:2006wn, GoldenfeldNigel1992Lopt}.  In this work we ignore the spatial correlations in environmental stochasticity because we wish to focus specifically on the role of short-range dispersal in generating large-scale synchrony. Correlated noise can be easily added to the dynamical Ising model and will be the subject of a subsequent study.  

\subsection{Overview}
\begin{figure}
    \centering
    \includegraphics[width = .7\textwidth]{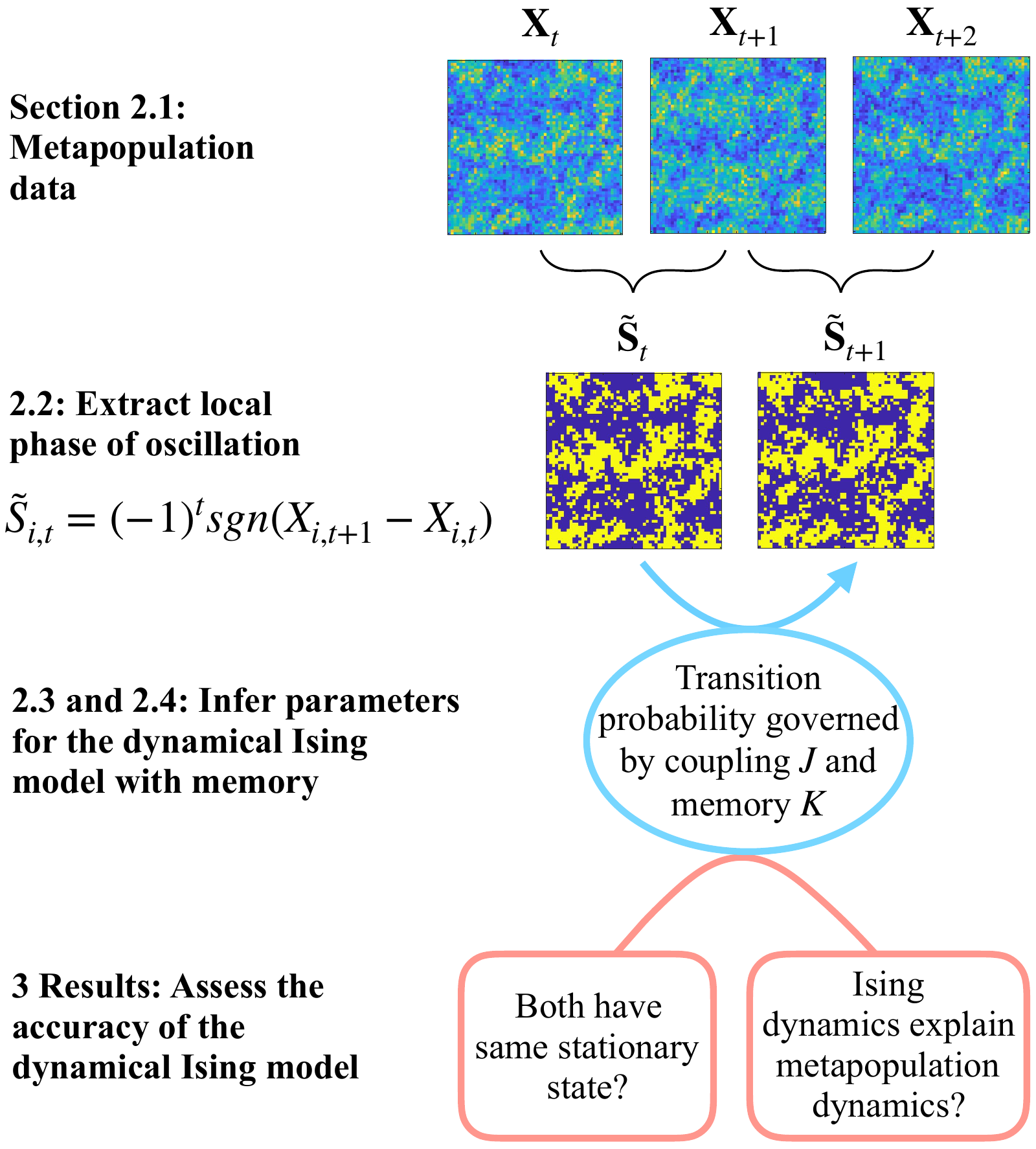}
    \caption{The steps in constructing and assessing a dynamical Ising model that best describes a two-cycle metapopulation. 
    }
    \label{flow}
\end{figure}

The overall program of the paper is sketched in Fig.\ \ref{flow}.  First, we carry out simulations of our metapopulation models, which consist of noisy, locally-coupled lattice maps (Sec.\ \ref{sec:ricker}).  The first row of the figure shows three successive snapshots of this model with the color code representing local subpopulation densities. From the (continuous) subpopulation densities we obtain a reduced description, where the phase of oscillation of each subpopulation is represented by a phase variable that takes values of $\pm 1$ (Sec.\ \ref{sec:synchrony}). A dynamical Ising model with memory introduced in Sec.\ \ref{sec:ising} equips the phase variables with stochastic, Markovian dynamics.  
Finally, from successive snapshots of the phase variables (second row of Fig.\ \ref{flow}) we use maximum likelihood inference methods (Appendix \ref{sec:inference}) to obtain values of the two parameters of the dynamical Ising model that best represent the dynamics of the metapopulation models.  Results from the application of this program and an assessment of the accuracy of the dynamical Ising representation  are presented in Sec.\ \ref{sec:results}. The stationary state and approach to the stationary state are studied in Sec.\ \ref{sec:resultJK} and \ref{sec:timedepend}, respectively, and the predictive power of Ising dynamics is tested in Sec.\ \ref{sec:flip}. The results show that the inferred dynamical Ising model is a reasonably good approximation to the more complex metapopulation dynamics for several metapopulation models and a wide range of parameters of the  models. Shortcomings of the dynamical Ising representation and an idea for how to improve it are discussed in Sec.\ \ref{sec:miss}.  The paper closes with a discussion in Sec.\ \ref{sec:discussion}.

\section{Models and methods}

\subsection{Dynamical metapopulation models: Noisy coupled lattice maps}
\label{sec:ricker}

The simplest metapopulation models  have identical local populations arrayed on a two-dimensional square lattice with only local dispersal, namely a noisy coupled lattice map \citep{HANSKI:1991vs, Durrett:1994ua}. Metapopulation models of this form with dispersal between four nearest neighbor patches on the lattice are studied on a grid of size $L \times L$ with $N=L^2$ patches. Let $X_{i,t}$ be the subpopulation density in patch $i$ at time $t$ and let $\mathbf{X}_t = \{X_{i,t}\}$ represent the collection of all subpopulation sizes.  We use periodic boundary conditions, which makes our lattice topologically equivalent to a torus.  While true ecological systems have edges, this set-up yields results that are more homogeneous and differ little from other possible boundary conditions except in small systems. 

Each subpopulation undergoes local dynamics,  interacts with its nearest neighbors through dispersal and is subject to uncorrelated environmental noise. These three processes happen in a sequence and are  represented, respectively, by three operators, $\mathbb{R}$, $\mathbb{D}$ and $\mathbb{N}$, described below.

\vspace{\baselineskip}
    \noindent\textbf{Local dynamics $\mathbb{R}$}:
    
   \noindent 
   Short-period oscillations in species with seasonal life histories and intraspecific density dependence are well described by discrete-time quadratic maps such as the Ricker and logistic models \citep{Fagan:2001vc, Murdoch:2002ii, Krkosk:2014ch, Abbottchapter}. 
   For example, the local dynamics with the Ricker map \citep{Ricker} acting on all the patches independently, $\mathbb{R}$, is given by
    \begin{equation}
        X_{i,t+1} \equiv (\mathbb{R}\mathbf{X}_t)_i = X_{i,t}\exp[r(1-X_{i,t})],
    \end{equation}
    whereas the local dynamics with the logistic map is given by
    \begin{equation}
        X_{i,t+1} \equiv (\mathbb{R}\mathbf{X}_t)_i = r X_{i,t} (1-X_{i,t}),
    \end{equation}
    where the parameter $r$ is the intrinsic population growth rate. Each patch in a metapopulation is assumed to undergo the same local dynamics and have the same growth rate.
    
    As $r$ increases, the behavior of the map goes through a series of bifurcations following the classic period doubling route to chaos \citep{MAY:1974uk}.  For the rest of the work, we choose $r$ such that the local dynamics is in the two-cycle regime so $X_{i,t}$ oscillates between high and low values.

\vspace{\baselineskip}
   \noindent\textbf{Dispersal $\mathbb{D}$}:

   \noindent  During dispersal a fraction $\epsilon$ of each subpopulation leaves its home patch and $\epsilon/4$ migrates to each of the four neighboring patches. The dispersal operator, $\mathbb{D}$, takes the form,
   \begin{equation}
            X_{i,t+1} \equiv (\mathbb{D} \mathbf{X}_t)_i = (1-\epsilon) X_{i,t} + \frac{\epsilon}{4} \sum_{\langle j;i \rangle} X_{j,t},
    \end{equation}
where $\langle j;i \rangle$ indicates a sum over the  nearest neighbors of site $i$. The values of dispersal $\epsilon \leq 1/2$ are reasonable from the biological standpoint by ensuring that a majority of the local population remains in its home patch.
  \vspace{2pt}

\vspace{\baselineskip}
\noindent \textbf{Noise $\mathbb{N}$}: 
 
\noindent We use log-normally distributed, multiplicative noise that acts independently on each patch at each time step, 
     \begin{equation}
         (\mathbb{N} \mathbf{X}_t)_i = X_{i,t} \exp(\lambda \zeta_{i,t})
     \end{equation}
where $\lambda$ is the noise level and $\zeta_{i,t}$ are independent, 
 identically distributed normal variates with mean zero and unit standard deviation. 
   
    \vspace{5pt}
     
  The three processes can be arranged in any order to form a metapopulation model. In the case of $\mathbb{NDR}\mathbf{X}_t$, each process acts once per cycle in the same order, first $\mathbb{R}$, then $\mathbb{D}$, and finally $\mathbb{N}$.  Any cyclic permutation of this order represents the same model but with the population censused at different stages in the cycle (see \citep{Ripa:2000vk} for a discussion of the effect of census time on measured synchrony).  On the other hand, $\mathbb{NRD}$ and its three cyclic permutations represent a distinct metapopulation model. We study four metapopulation models with different choices of local dynamics and sequence of processes (Table \ref{tab:four_models}).
  For a given model, the additional three parameters are the number of patches, $N$, dispersal, $\epsilon$, and noise, $\lambda$.

\begin{table}[]
    \centering
    \begin{tabular}{|c|c|c|c|}
    \hline
         Model & Local map & Growth parameter & Sequence  \\
         \hline
        A & Ricker & $r=2.2$ & $\mathbb{NDR}$ \\
        B & Ricker & $r=2.4$ & $\mathbb{NDR}$ \\
        C & Logistic & $r=3.2$ & $\mathbb{NDR}$ \\
        D & Ricker & $r=2.2$ & $\mathbb{NRD}$ \\
        \hline
    \end{tabular}
    \caption{Metapopulation models studied here. The models are defined in detail in Appendix \ref{sec:rickerNDR}.  Except where otherwise state, reported results refer to Model A.  In all cases, measurement is done after the action of noise. In Model D, the order of local dynamics and dispersal is reversed.
    }
    \label{tab:four_models}
\end{table}
       
       \subsection{Synchrony and Phase Transitions in Metapopulation Models}
       \label{sec:synchrony}
A discrete two-cycle oscillator, for example the Ricker map at $r=2.2$, has two possible phases of oscillation. In what we define arbitrarily as the ``plus'' phase, the high value of the oscillator occurs at odd times and in the ``minus'' phase, the high values occur at even times. We define a two-cycle variable, $M_{i,t}$, for each subpopulation $i$ at time $t$ as an alternating-sign first difference,
\begin{equation}\label{two-cycleamplitude}
           M_{i,t} = \frac{1}{2} (-1)^{t}(X_{i,t+1}-X_{i,t}).
          \end{equation}
When $M_{i,t}$ is positive (negative) the oscillation in patch $i$ is in the plus (minus) phase.
       
\begin{figure}
    \centering
    \includegraphics[width=0.8\textwidth]{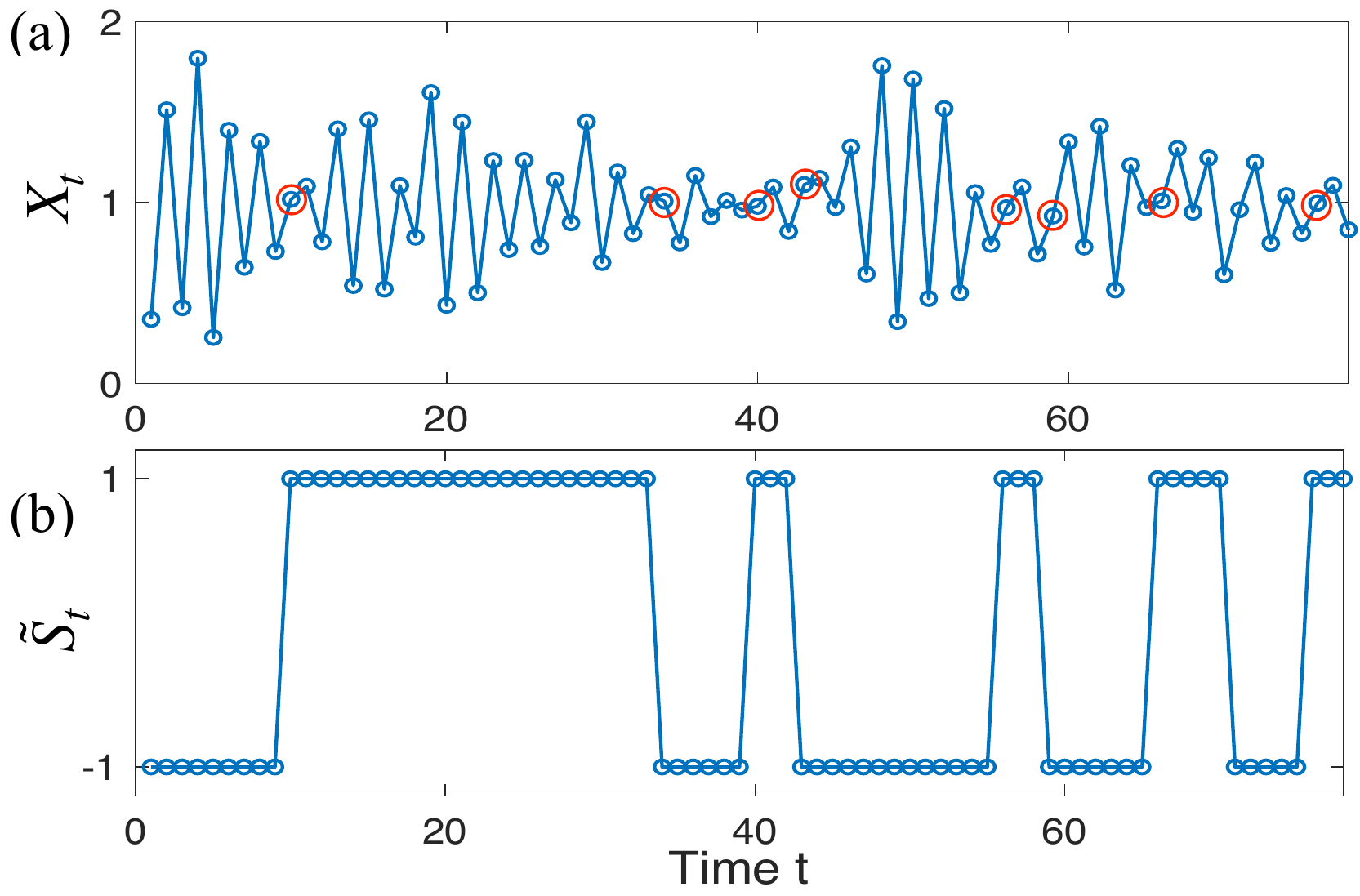}
    \caption{(a)  A time series of the Ricker variable $X_t$ for a single noisy Ricker map in the two-cycle regime ($r=2.2$) with noise level, $\lambda = 0.15$. The amplitude and phase of the oscillations vary due to the noise.  Times when the phase of the oscillation changes are marked with red circles. (b) The time series of  $\tilde{S}_t$, representing the phase of oscillation of the Ricker variable, is obtained from the time series $X_t$  (Eq.\ \eqref{mapping}). 
    }
    \label{flip}
\end{figure}

The variable $M_{i,t}$ contains both amplitude and phase information.  We can extract the phase, $\tilde{S}_{i,t}$, of the oscillation by taking the sign (signum function, sgn) of $M_{i,t}$, 
       \begin{equation}\label{mapping}
             \tilde{S}_{i,t} =  \mathrm{sgn}(M_{i,t})
       \end{equation}
where $\tilde{S}_{i,t}$ takes  values $\pm 1$.  We refer to $\tilde{S}_{i,t}$ as the subpopulation phase variable.
For a single local oscillator without noise in a steady state two-cycle, the phase variable is independent of time. In the presence of noise, the phase of oscillation and thus the phase variable changes stochastically, as shown in Fig.\ \ref{flip}, where the points circled in red denote times when the phase of oscillation changes.

Let $\mathbf{M}_{t}=\{M_{i,t}\}$ and $\mathbf{\tilde{S}}_{t}=\{\tilde{S}_{i,t}\}$ denote the collection of two-cycle variables and associated phase variables, respectively, for all subpopulations.

For much of the remainder of this study, we consider the behavior of the metapopulation model after it has reached a stationary state where the statistical properties of $\mathbf{M}_t$ and, therefore, $\mathbf{\tilde{S}}_t$ are time-independent.
The spatial and temporal average of $\tilde{S}_{i,t}$ or, equivalently, the average over the stationary distribution \citep{Durrett:1999},  $\order$,
\begin{equation} \label{eq:littles}
        \order(\lambda,\epsilon)  = \frac{1}{T-T_b} \frac{1}{ N} \sum_{t=T_b}^{T-1} \Big\vert \sum _{i=1}^N \tilde{S}_{i,t} \Big\vert\ ,
\end{equation}
is called the  ``synchronization order parameter" and measures the degree of synchrony in the entire metapopulation. $T_b$ is a time we wait for the system to reach  its stationary state and $T$ a much longer time. For a large system, the metapopulation may remain synchronized for very long times even though the  phase variable of individual oscillators may change on relatively short time scales.

\begin{figure}
            \centering
        \includegraphics[width = .99\textwidth]{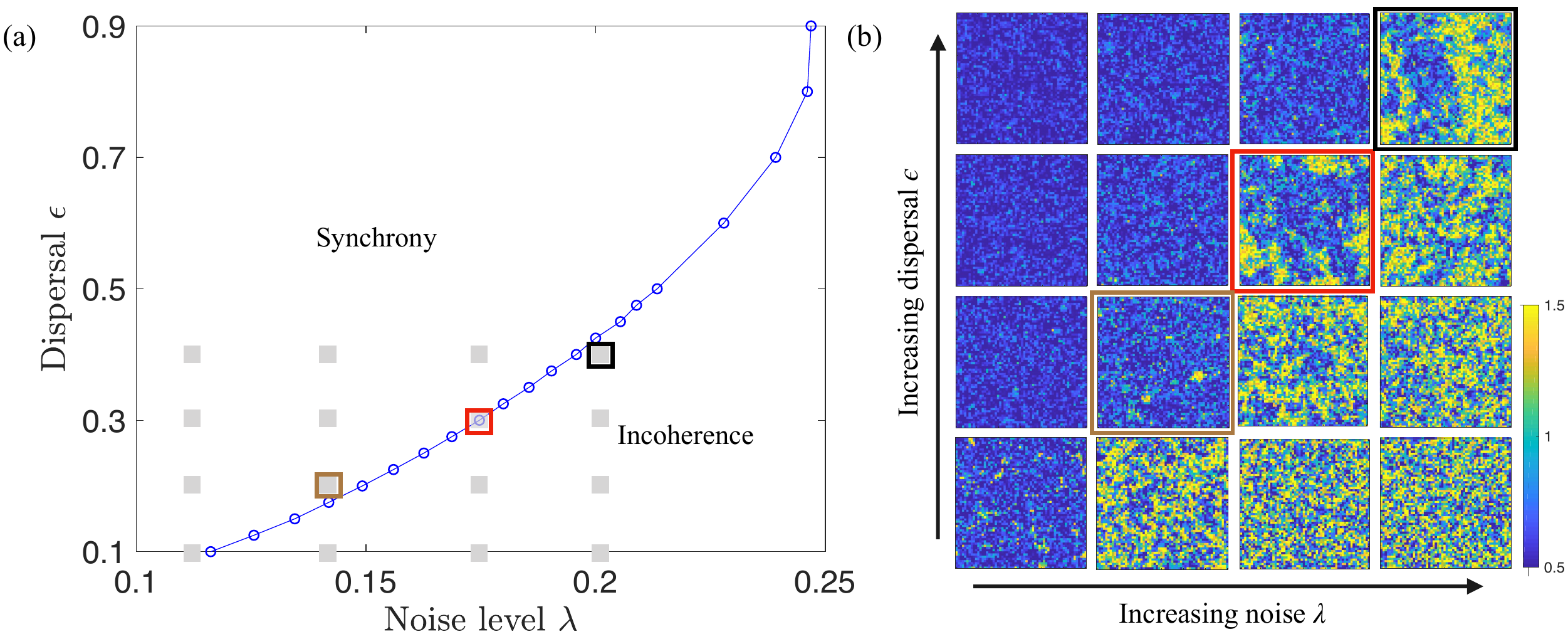}
        \caption{(a) The state diagram for  metapopulation Model A showing the synchronous (ordered) and incoherent (disordered) regions  in the noise-dispersal plane. The two regions are separated by a critical transition  (blue line).  The open circles show the values of dispersal $\epsilon$ for which simulations were performed to find the critical values of noise $\lambda_c(\epsilon)$ 
        and the blue line connects the points. 
        (b) Typical snapshots showing local population densities in each patch of a $60\times 60$ metapopulation. The dispersal and noise value for each snapshot is shown as a corresponding gray square on the state diagram. The snapshot outlined in red is on the critical line and shows long-range correlations. The snapshots outlined in brown and black are about same distance from critical line but the snapshot in incoherent state (outlined in black) looks more critical. The snapshots near the upper left are in the synchronous state and display global synchrony while the snapshots near the lower right are in the incoherent state and subpopulations are uncorrelated. 
        }
        \label{phase plot}
        \end{figure}

The metapopulation's stationary state depends on the noise and dispersal parameters and, as a function of these parameters, undergoes a critical transition (also sometimes referred to as a continuous or second-order phase transition)
 in the following sense.  The synchronization order parameter, $\order(\lambda,\epsilon)$, changes continuously as a function of both noise $\lambda$ and dispersal $\epsilon$.  For a given value of dispersal and noise less than a critical value, $\lambda_c(\epsilon)$, the synchronization order parameter is nonzero and its magnitude approaches one as $\lambda$ approaches zero.   As $\lambda$ increases for fixed $\epsilon$, the synchronization order parameter decreases and it is very near zero for all values $\lambda \geq \lambda_c(\epsilon)$.  Figure \ref{phase plot}a is the state diagram (also referred to as phase diagram) obtained for a metapopulation Model A showing regions in the noise-dispersal ($\lambda$--$\epsilon$) plane where there is  synchronous (ordered, $|\order|>0$) and incoherent (disordered, $\order \approx 0$) behavior together with the critical line, $\lambda_c(\epsilon)$, separating these regions.  As expected, for higher values of dispersal, synchrony is maintained for higher values of noise.  To obtain the critical line, the critical noise, $\lambda_c(\epsilon)$, is found for a number of $\epsilon$ values using the Binder cumulant method (Appendix \ref{sec:Binder}) \citep{Binder1981}.

 The  transition from incoherence to synchrony in the metapopulation model is a critical transition because it exhibits two related features: large fluctuations and long-ranged correlations.  These features can be seen qualitatively in Fig.\ \ref{phase plot}b, which shows typical snapshots of the local population variables of the metapopulation after it has reached the stationary state for different values of dispersal and noise. 
As shown in \citep{Noble2015}, the critical transition of many noisy lattice map systems in the two-cycle regime, including the metapopulation models studied here, are in the Ising universality class.

\subsection{Dynamical Ising model with memory}
\label{sec:ising}

In this work we seek to go beyond universal properties and understand whether the Ising model can provide a good description of the local dynamics of a metapopulation\footnote{There are universal dynamical properties but these apply to asymptotic long-time and long-distance properties of time correlation functions not the local and short time properties studied here.}.  For this purpose we need to introduce dynamical Ising models.  Several dynamical Ising models have been studied in statistical physics~\citep{Hohenberg:1977ym}. All of these models have stationary state properties described by a Gibbs distributions in the Ising universality class but they differ in their dynamics.  The standard dynamical model, sometimes known as heat bath dynamics or the Glauber model \citep{glauber1963, newmanb1999}, has no self-interaction.  For this reason, it fails to capture an essential feature of the metapopulation model -- that each subpopulation tends to preserve its own phase of oscillation, as is seen in Fig.\ \ref{flip}.  In this section and   Appendix \ref{app:ising} we introduce the dynamical Ising model with memory, which provides a far better description of the metapopulation dynamics.  

The degrees of freedom of 
Ising models are Ising spins\footnote{We use the notation $\mathbf{S}_t$ to refer to the spin configuration of a dynamical Ising model, in contrast to $\tilde{\mathbf{S}}_t$  that represents the subpopulation phases of oscillation as described in Sec. \ref{sec:synchrony}. 
}, $\mathbf{S}_t= \{S_{i,t}\}$ with $S_{i,t}=\pm 1$.
In the standard dynamical Ising model the state of each spin at time $t+1$ is influenced by the state at time $t$ of its nearest neighbors with an interaction strength, $J$.  For the dynamical Ising model with memory \citep{CIRILLO201436, Pre}, the state of a spin at time $t+1$ is additionally influenced by its own state at time $t$  with a self-interaction strength, $K$. The spatial structure and thus the neighborhood of each spin in the dynamical Ising model is the same square grid as in the metapopulation models. Spin are updated in parallel as they are in the metapopulation models.

The dynamics of the Ising model with memory is most succinctly stated in term of the probability, $P(S_{i,t+1}=-S_{i,t}| \mathbf{S}_{t},J,K)$,  that spin $S_{i,t}$ changes sign from time $t$ to $t+1$.  This ``flip probability" depends only on the state of the spin and it four nearest neighbors according the flip probability function $P_f$,
\begin{equation}\label{eqn:flip}
    P(S_{i,t+1}=-S_{i,t}| \mathbf{S}_{t},J,K) =  P_f(h_{i,t}S_{i,t}),
\end{equation}
where $h_{i,t}$ is the sum of the neighbor spins,
\begin{equation}
\label{eq:h}
    h_{i,t} = \sum_{\langle j;i \rangle} S_{j,t},
\end{equation}
and the flip probability function takes the form,
\begin{equation}
\label{eq:pf}
 P_f(x) =  \frac{\exp\big(-J x- K\big)}{2\cosh\big(J x+K \big)}.   
\end{equation}
Flips are suppressed when a spin is in the same state as the majority of its neighbors $x=h_{i,t}S_{i,t} >0$ and vice versa.  When the memory term $K$ is large, flips are suppressed regardless of the state of the neighbors,  $P_f \rightarrow e^{-2K}$, and the dynamics of the system is very slow.

\subsection{Ising representation of the  metapopulation models}

In this section we describe the methods we use to obtain the best dynamical Ising model representation of a metapopulation and then to assess the fidelity of that representation.  For a given metapopulation model, we carry out simulations to generate a large set of successive triples of snapshots of the subpopulation densities, $(\mathbf{X}_{t},\mathbf{X}_{t+1},\mathbf{X}_{t+2})$ for various values of dispersal and noise along its critical line, $\lambda_c(\epsilon)$.  From each triple,  we obtain a pair of phase configurations, $(\mathbf{\tilde{S}}_{t},\mathbf{\tilde{S}}_{t+1})$ (Fig.~\ref{flow}). We use these generated pairs to infer the parameters $J$ and $K$ of the dynamical Ising model that maximize the likelihood of producing the same distribution of pairs of Ising spin configurations $(\mathbf{S}_{t},\mathbf{S}_{t+1})$.  We describe the dynamical inference method in Appendix \ref{sec:inference}. Note that the inference method is based solely on the flip probabilities and hence can be applied to  metapopulation data at any time, not just in the stationary state. Furthermore, since the phase variables and Ising variables are binary, the inference procedure is straightforward to implement and reasonably accurate even with relatively little metapopulation data. 

\begin{figure}
    \centering
    \includegraphics[width = .5\textwidth]{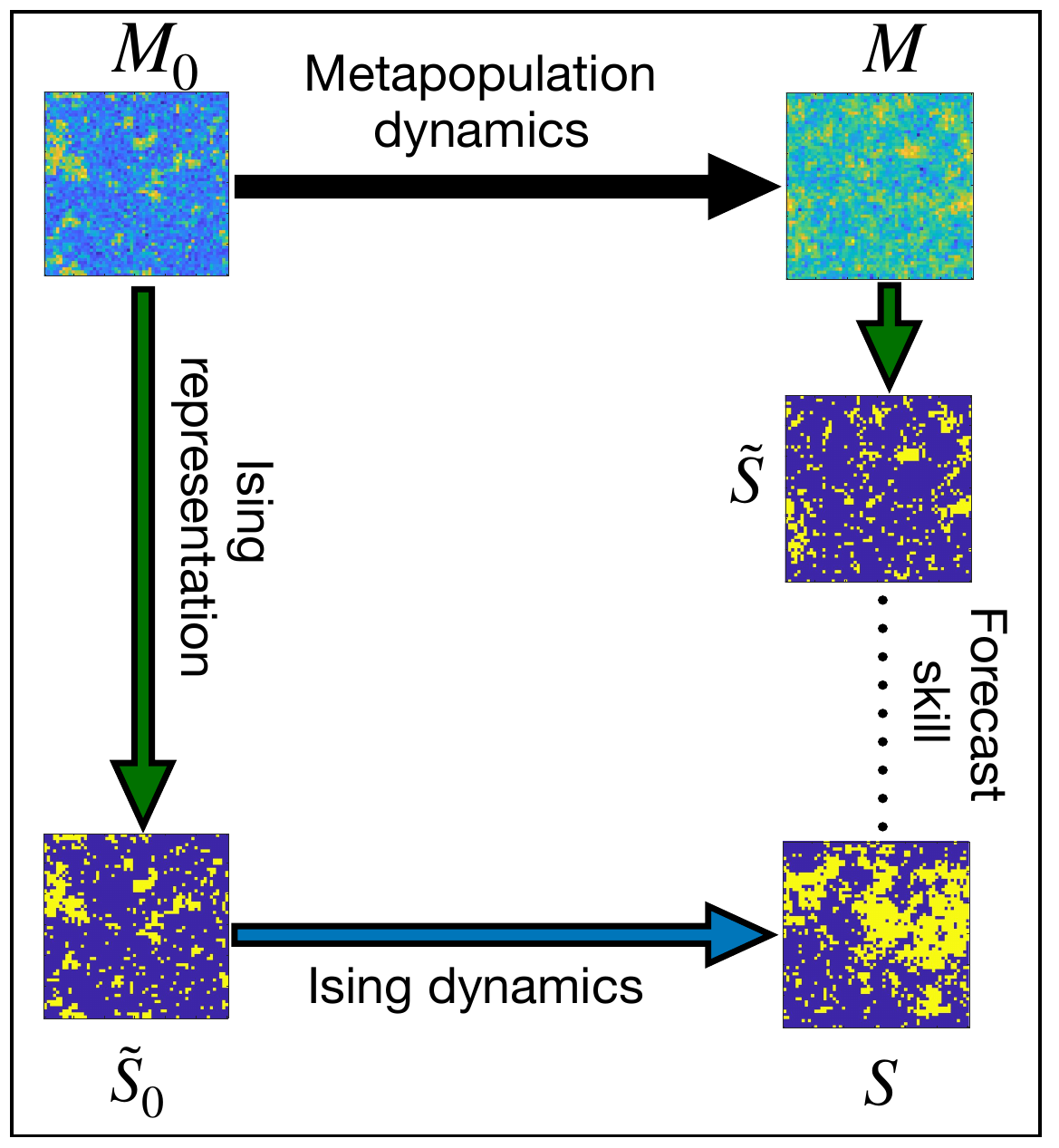}
    \caption{Using Ising dynamics to predict metapopulation dynamics.  
    The initial state is a metapopulation two-cycle configuration $\mathbf{M}_0$ chosen from the stationary distribution and $\mathbf{M}$ is the configuration at the next time step. The input for the Ising model prediction is the configuration of phase variables, $\tilde{\mathbf{S}}_0$ obtained from $\mathbf{M}_0$. The prediction of the subsequent configuration, $\mathbf{S}$ is compared to the actual configuration $\tilde{\mathbf{S}}$ using a forecast skill score.   
    }
    
   \label{fig:FS_MI}
\end{figure}

We can use the inferred Ising model as a predictive tool as follows:  Suppose we are given information about the metapopulation at times $t$ and $t+1$  in the form of the phase variables $\mathbf{\tilde{S}}_{t}$ (which involves the transition between times $t$ and $t+1$).  We can use these data as an initial value for the dynamical Ising model by setting $\mathbf{S}_{t}=\mathbf{\tilde{S}}_{t}$, and  then calculate the  flip probability given by Eq.~\eqref{eqn:flip} to predict the future state of the metapopulation from the Ising model. These probabilistic predictions are then compared to the metapopulation data and assessed using measures of forecast skill.  The setup is sketched in Fig.\ \ref{fig:FS_MI}.

\section{Results}

In this section we present the results of our metapopulation simulations and their Ising representations. Our objective is to provide a broad assessment of the dynamical Ising model with memory as an effective tool for describing and predicting metapopulation dynamics.  In Sec.\ \ref{sec:resultJK} we test whether the inferred Ising model has stationary states which are close to the critical stationary states of the metapopulation models they are representing.  
In Sec.\ \ref{sec:timedepend} we explore how the best fit Ising parameters change as a function of time as the metapopulation goes from an initially random configuration toward the stationary state.  
In Sec.\ \ref{sec:flip} we compare the flip probability  of an Ising spin with the probability of a phase change in a subpopulation
and measure the forecast skill of the dynamical Ising model to predict phase changes in subpopulations.  
Finally, in Sec.\ \ref{sec:miss} we identify what the dynamical Ising model misses that cause errors in modeling the metapopulation.
\label{sec:results}

\subsection{Inferred Ising parameters in the stationary state}
\label{sec:resultJK}
\begin{figure}
    \centering
    \includegraphics[width = .99\textwidth]{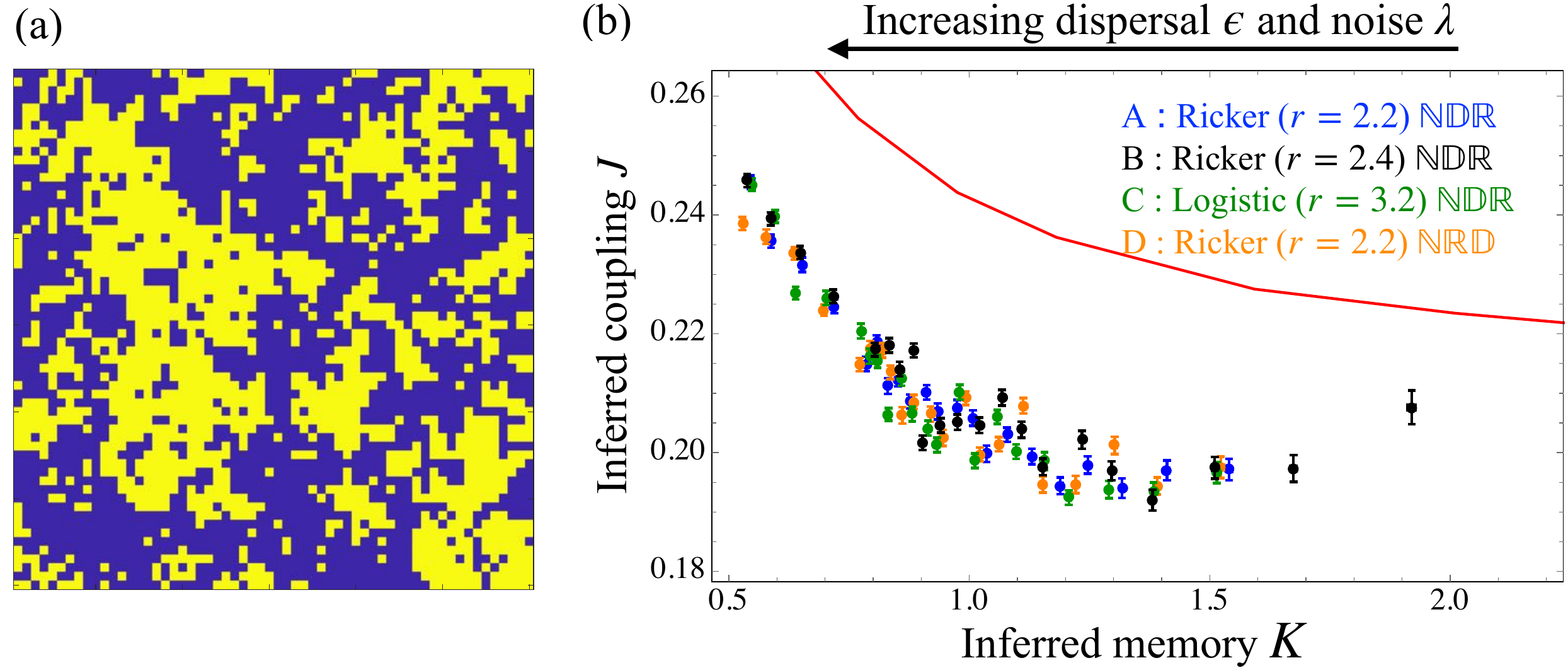}
    \caption{(a) A typical spin configuration in the stationary state of the Ising model with memory simulated with the best fit parameters inferred from  metapopulation Model A with parameters $\epsilon = 0.325$ and critical noise $\noise = 0.179$.  (b) Inferred Ising parameters are plotted in the coupling-memory ($J$--$K$) plane and are compared with critical curve (red line) of the dynamical Ising model with memory.  Each inferred ($J$,$K$) point corresponds to a simulation point ($\lambda=\noise$) on the critical line of the corresponding metapopulation model (see Figs.\ \ref{phase plot} and \ref{fig:phase_AD}), with increasing $K$ corresponding to decreasing noise. }
    \label{phasetwo}
\end{figure}

Figure  \ref{phasetwo}a shows a snapshot of the Ising model with memory in the stationary state simulated with the inferred parameters for the metapopulation Model A with dispersal $\epsilon = 0.325$ and critical noise
$\noise = 0.179$. The snapshot displays long range correlations and is visually similar to the corresponding critical snapshot of the metapopulation model (Fig.\ \ref{phase plot}b, red box).  
 
Figure \ref{phasetwo}b shows the maximum likelihood parameters $J$ and $K$ of the dynamical Ising model inferred from simulation data of the four metapopulation models for various values of dispersal $\epsilon$ and noise $\noise$. Each colored point represents the inferred value of $J$ and $K$ for a given value of  dispersal $\epsilon$ and  noise $\noise$ chosen from the critical line of the appropriate metapopulation model. For Model A, the critical line is shown in Fig.\ \ref{phase plot} (and  Fig.~\ref{fig:phase_AD} for other models).  The order of points in Fig.\ \ref{phasetwo}b are the reverse of the order of the points in Fig.\ \ref{phase plot}. This ordering can be understood intuitively from the observation that increasing noise should reduce the stability of the phase of the local oscillators and increasing dispersal should increase the coupling between neighboring oscillators.  Thus we expect that with increasing noise and dispersal, the inferred memory $K$ will decrease and the inferred coupling $J$ will increase, as clearly seen in Fig.\ \ref{phasetwo}b.

The critical line of the dynamical Ising model with memory from Ref. \citep{Pre}
is re-plotted in red in Fig.~\ref{phasetwo}b. Since the stationary state of the metapopulation models is at a critical point for each of the simulated values of dispersal and noise, a perfectly accurate Ising representation of the metapopulation would also have critical stationary states.  We see that this is not quite the case, and instead the inferred Ising models have stationary states close to but clearly below the critical line in the disordered state.  It is perhaps not surprising that a simplified model that is inferred from the local dynamical properties of a metapopulation fails to exactly capture the large scale ordering properties of the system in the stationary state. We explore a possible cause of this failure in Sec.~\ref{sec:miss}.  Nonetheless, the inferred parameters are close enough to the Ising critical line to display long-ranged correlations (see Fig.\ \ref{phasetwo}a).

Figure~\ref{phasetwo}b also shows that the dynamical Ising  representation is robust and performs very similarly independent of the underlying metapopulation dynamics since the results for the four different metapopulation models all lie on nearly the same curve.

\subsection{Time dependence of inferred Ising parameters}
\label{sec:timedepend}

\begin{figure}
    \centering
    \includegraphics[width = .6\textwidth]{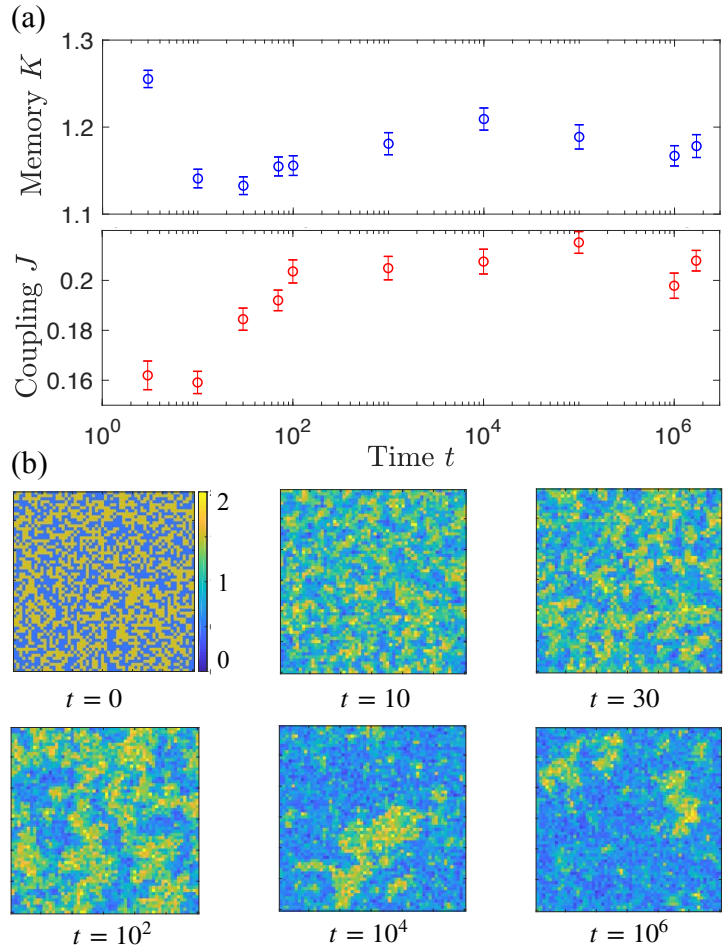}
    \caption{(a) Inferred parameters $J$ and $K$ as a function of time for metapopulation Model A with random initial  conditions and with $\epsilon = 0.2$ and $\lambda_c(\epsilon=0.2)=0.15$. At $t=0$ each subpopulation is equally likely to be in the high or low  state of the two-cycle Ricker map. The inferred values at late times  match the stationary state results in Fig.\ \ref{phasetwo}. (b) Lattice snapshots of the metapopulation model at several times show the evolution of the system from the random initial configuration to the stationary state.}
    \label{timeplot}
\end{figure}

The inferred Ising results shown in Fig.\ \ref{phasetwo}b are obtained from simulation results after the metapopulation has reached the stationary state. We repeat the inference calculation at various times before the metapopulation has reached its stationary state for the case $\epsilon=0.2$ and its critical noise $\noise=0.15$, starting from a random initial condition with half of the subpopulations at the low value of the two-cycle and the other half at the high value (Fig.\ \ref{timeplot}b, $t=0$).

Figure \ref{timeplot}a shows the inferred parameters at different times after random initial conditions. The snapshots in Fig.\ \ref{timeplot}b show the evolution of the configurations with time $t$.  Long-range correlations and  critical behavior develop slowly and are not fully visible until $t=10^4$.
We observe that at early times, the inferred values of $J$ and $K$ decrease and later increase to saturate at the inferred values in the stationary state. Interestingly, this saturation occurs earlier ($t \approx 10^2$) than when the critical pattern characteristic of the stationary state first appears ($t \approx 10^4$).
A perfectly accurate dynamical model should have parameters that are independent of the current state of the system.  The relatively weak time dependence of the parameters is another indication that the dynamical Ising model is a good but not perfect approximation to the metapopulation model.

\subsection{Flip probability and forecast skill}
\label{sec:flip}

\begin{figure}
    \centering
    \includegraphics[width = .9\textwidth]{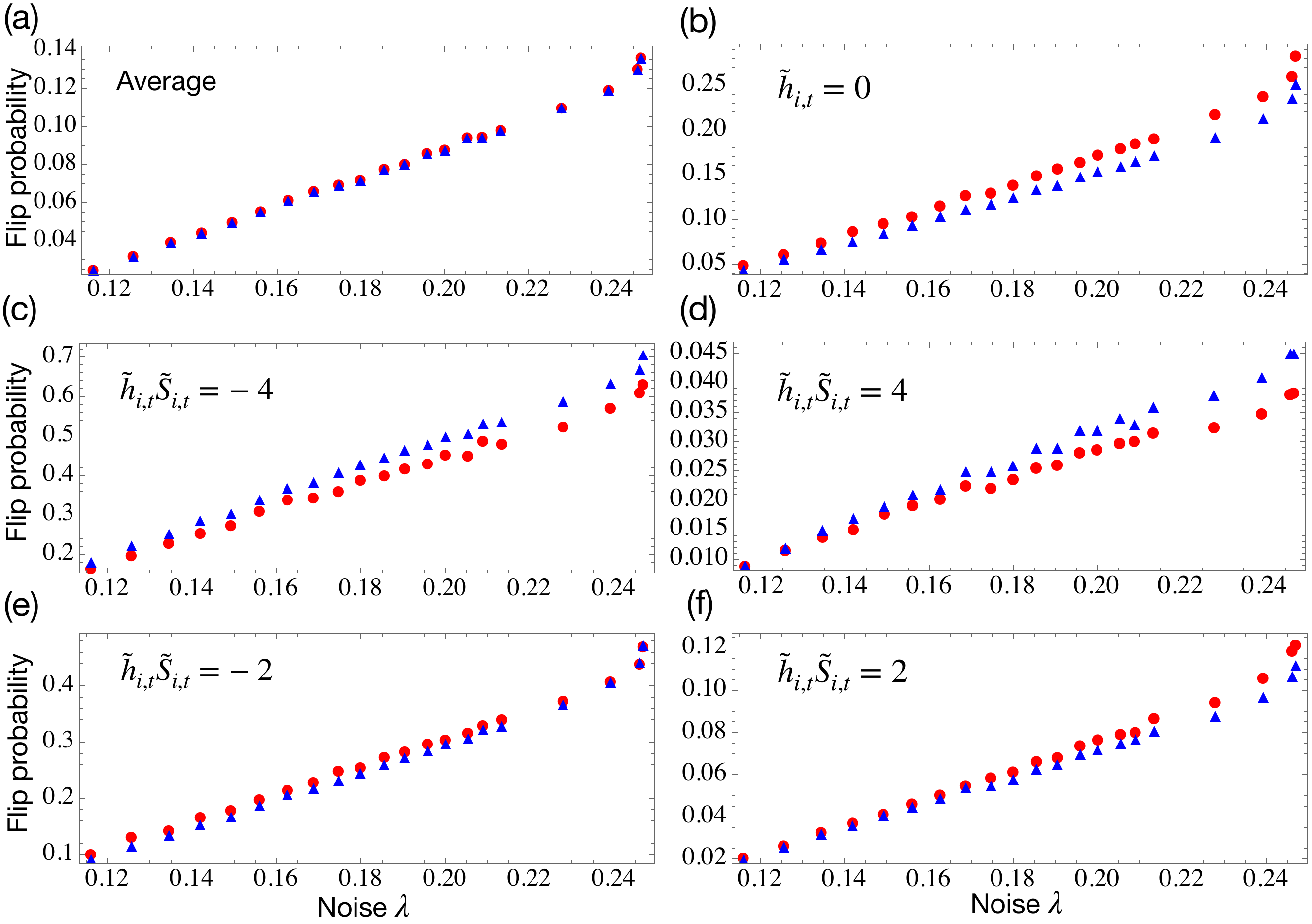}
    \caption{Probability of a phase change (``flip'') for metapopulation Model A (red circles) and predictions of the Ising model with memory (blue triangles). Panel (a) shows the average flip probability in the stationary state.  The remaining panels show flip probabilities conditioned on the possible values of the product $\tilde{h}_{i,t} \tilde{S}_{i,t}$. When this product is positive(negative), the phase of a subpopulation $i$ at time $t$ agrees(disagrees) with the majority of its neighbors.  When $\tilde{h}_{i,t}=0$ the influence of the neighbors is weak. 
    }
     \label{noneighbor}
    
\end{figure}

Figure \ref{noneighbor}a shows the average flip probablility (rate of phase changes of subpopulations) measured in the steady state of the  metapopulation  compared to the prediction of the inferred dynamical Ising model, $\mathbb{E}[P_f(\tilde{h}_i \tilde{S}_i)]$,   where the expectation is over values of  $\tilde{h}_i \tilde{S}_i$ in the metapopulation steady state and $P_f$ is defined in Eq.\ \eqref{eqn:flip}.  The Ising model very accurately predicts the average rate of phase changes.  

The remaining panels of Fig.\ \ref{noneighbor} show flip probabilities conditioned on specific initial (time $t$) values of the product $\tilde{h}_i \tilde{S}_i$, which take values $\pm 4$, $\pm 2$ and 0. When $\tilde{h}_i \tilde{S}_i$  is  positive, the neighbors of subpopulation $i$ have the same phase of oscillation as subpopulation $i$ and phase changes are discouraged while when $\tilde{h}_i \tilde{S}_i$ is negative, phase changes are encouraged, as evidenced in the plots.  Again, the predictions of the dynamical Ising model with memory, $P_f(\tilde{h}_i \tilde{S}_i)$ are quite accurate.  These results show that the effects of neighbors are very important for understanding the behavior of a subpopulation and that the Ising model does a good job of capturing these effects through the neighbor coupling, $J$. The small value of the flip probability, especially for weak noise shows, the importance of phase memory, captured in the memory parameter $K$. Note that the standard ($K=0$) dynamical Ising model predicts a flip probability of $1/2$ for  $\tilde{h}_i = 0$, which is far from the observed behavior of the metapopulation model (Figure \ref{noneighbor}b).  It is perhaps not surprising that the Ising model with memory does a quite good job at predicting the five conditional flip probabilities shown in Fig.\ \ref{noneighbor}(b-f) since these five measured numbers are the inputs to the dynamical inference procedure that yields $J$ and $K$ (see Appendix \ref{sec:inference}).

\begin{figure}
    \centering
    \includegraphics[width = 0.7\textwidth]{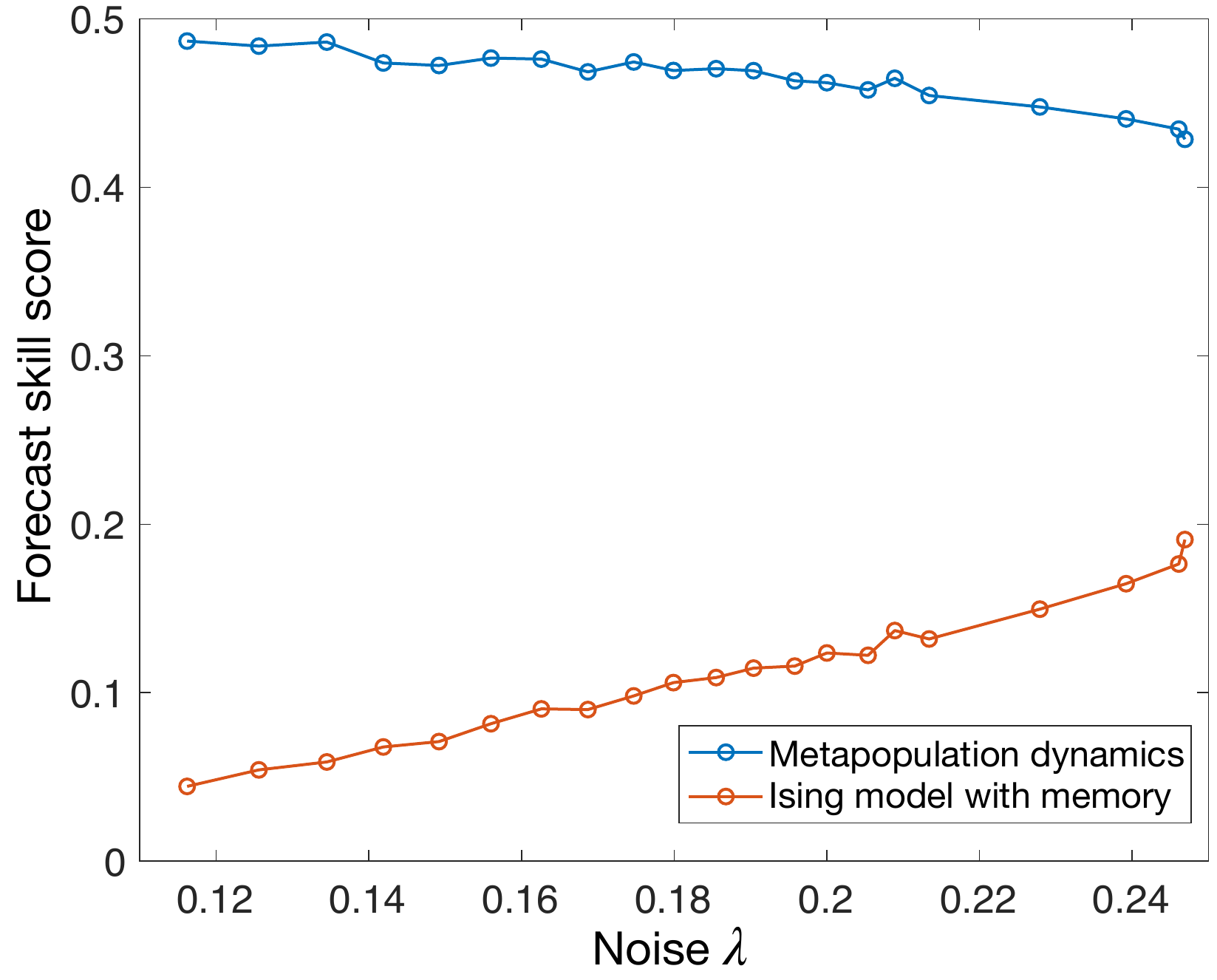}
    \caption{Forecast skill score for predicting phase flips is measured as a function of noise strength $\noise$ for values of noise and dispersal along the critical curve of the metapopulation model.  The reference forecast is the average flip probability.  The forecast skill of the metapopulation model itself (top, blue line) is less than 1 due to stochasticity and serves as an upper bound on predictability. 
    }
    \label{forecastskillplot}
\end{figure}

A more stringent test of the predictive power of the Ising model is the forecast skill in predicting whether a specific subpopulation $i$ will undergo a phase change from time $t$ to $t+1$ given the time $t$ data, $\tilde{h}_{i,t} \tilde{S}_{i,t}$.  The Ising prediction for the probability of this flip is $P_f(\tilde{h}_i \tilde{S}_i)$.  Figure \ref{forecastskillplot} shows the Brier forecast skill score (Appendix \ref{sec:forecastskill}) for this probabilistic prediction.  The forecast skill score compares the forecast to a reference forecast, which we here choose as the overall average rate of phase changes (red points in Fig.\ \ref{noneighbor}a). The forecast skill score is bounded by the skill score of the underlying metapopulation model, shown as the upper curve in Fig.\ \ref{forecastskillplot}, which is less than one and decreasing with noise because the metapopulation model is inherently stochastic.  The forecast skill of the Ising model with memory is relatively far from the bound, especially for low noise.  We believe this lack of skill is the result of the loss of information going from the original continuous two-cycle variables $M_{i,t}$ to the binary phase variables $\tilde{S}_{i,t}$ as discussed in the next section.

\subsection{What does the dynamical Ising model miss?}
\label{sec:miss}
\begin{figure}
    \centering
    \includegraphics[width = 1\textwidth]{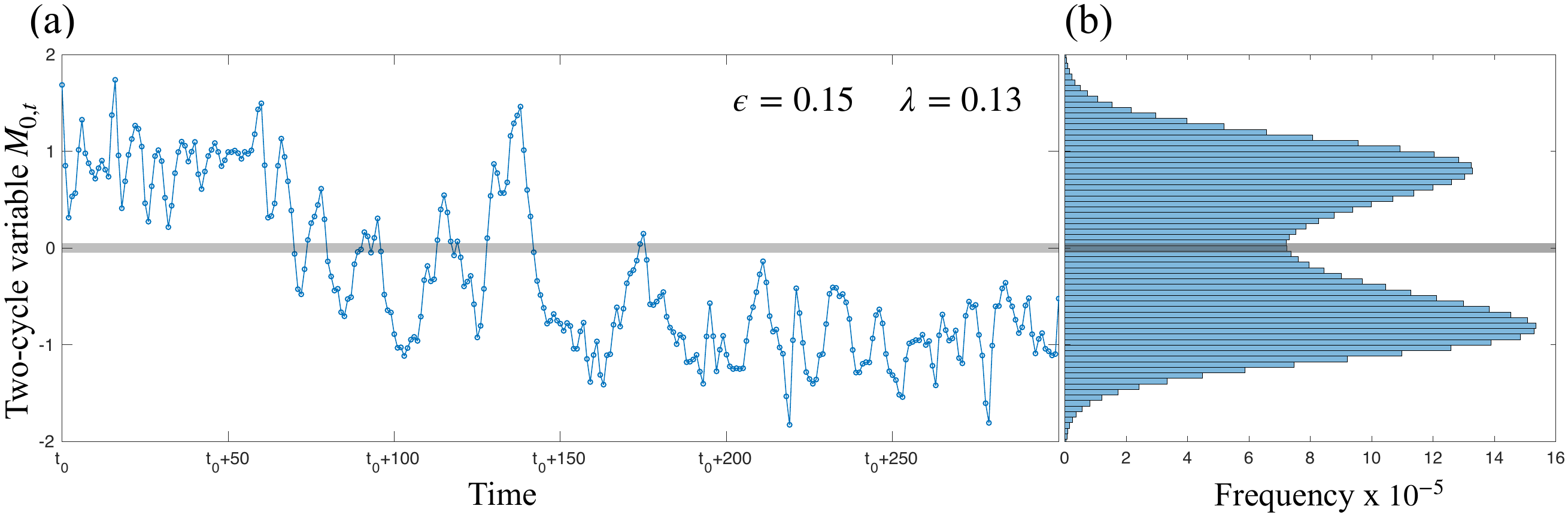}
    \caption{(a) A typical time series of the two-cycle variable $M_{0,t}$
    for a single subpopulation in metapopulation model A with dispersal $\epsilon=0.15$ and noise $\noise = 0.13$. The time $t_0$ is after the metapopulation model has reached the steady state. The region shaded gray highlights  times when the oscillation amplitude is near zero.  Changes in the phase of oscillation tend to occur when the amplitude has a value near zero.
    (b) Distribution of the subpopulation two-cycle variable, $M_{0,t}$.  
    }
    \label{time}
\end{figure}

In the previous section we saw that the dynamical Ising model with memory displays low skill in predicting phase changes in the metapopulation model for weak noise.  We believe that this deficiency is the result of using a  model with binary variables to predict the behavior of a system with continuous variables.  Figure \ref{time}a shows a time series of the two-cycle variable of $M_{0,t}$ (see Eq.\ \eqref{two-cycleamplitude}) of subpopulation $0$ in the steady state.  The two-cycle variable contains both phase and amplitude information.  Consider times in the figure where the phase variable changes sign ($\tilde{S}_{0,t+1}=-\tilde{S}_{0,t}$).  These events invariably occur when the amplitude of $M_{0,t}$ is small, with a value in or near the narrow grey band in the figure.   On the other hand, the binary variables $S_{i,t}$ contain no amplitude information.  Dynamical Ising models cannot capture the fact that phase changes occur when the amplitude of oscillation becomes small. We believe this defect explains why the forecast skill score is relatively low and also why the inferred Ising parameters do not fall exactly on the critical curve of the Ising model with memory.

\begin{figure}
    \centering
    \includegraphics[width = 0.7\textwidth]{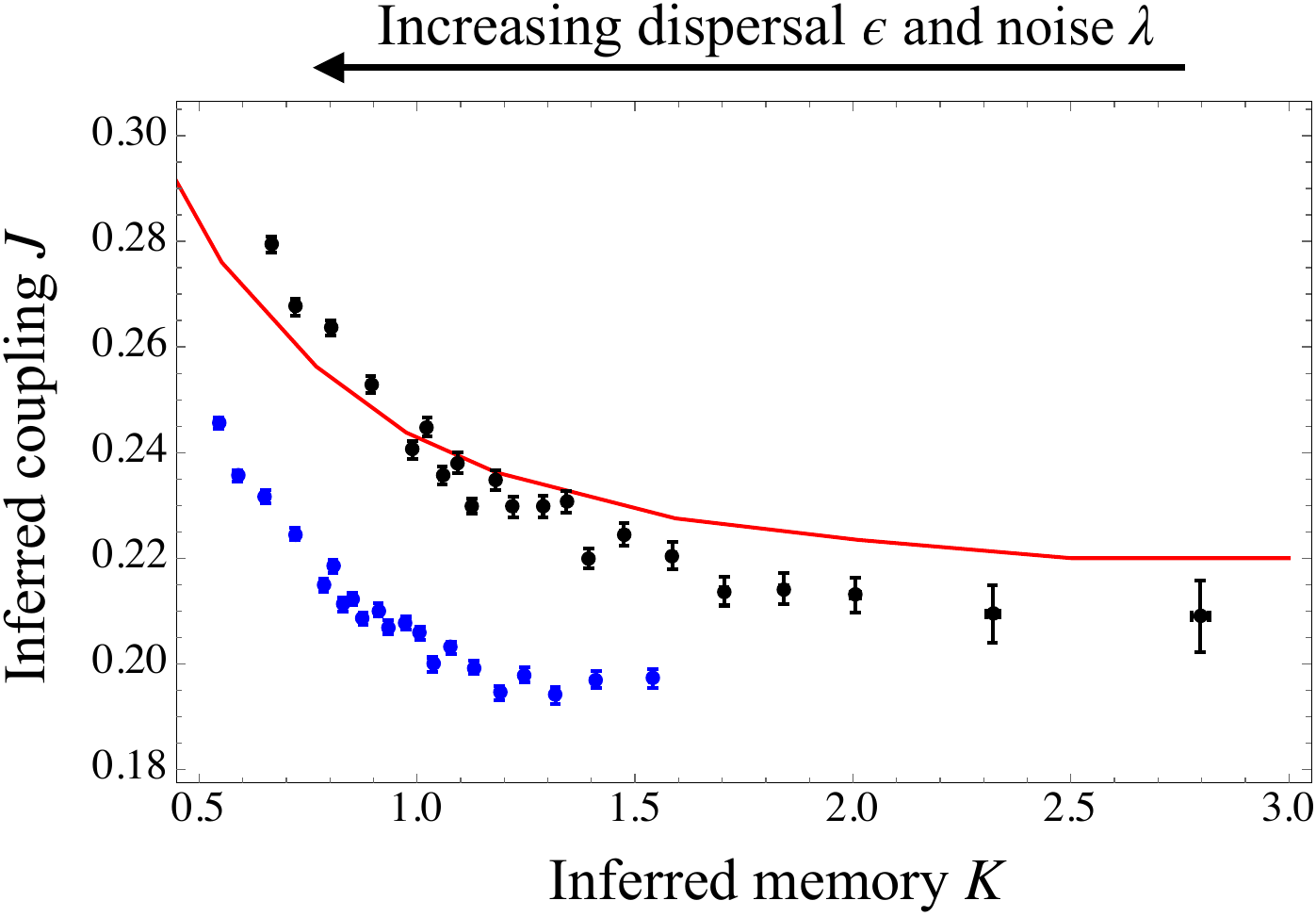}
    \caption{Inference results using the full dataset (blue, same as Fig.\ \ref{phase plot}) and using a subset of the data (black) obtained by discarding the $5\%$ smallest absolute values of the two-cycle variable $M_{i,t}$ (shaded region in Fig.\ \ref{time}a). 
    } 
    \label{blumeplot}
\end{figure}

To test the latter hypothesis, we discarded the 5\% of the subpopulations in the metapopulation data with the smallest values of $M_{i,t}$ (gray band in Fig.~\ref{time}) and reran inference on the remaining data.  The results are shown in Fig.~\ref{blumeplot}. The parameters $J$ and $K$ now fall much closer to the critical line of the Ising model with memory. This success suggests using a generalization of the Ising model with three states, $-1$, $0$ and $+1$ to better capture the fact that phase changes typically occur when the amplitude of oscillation is near zero.  The Blume-Capel model \citep{Blume66, Capel66} is an appropriate three-state generalization of the Ising model that will be the subject of a future study. 

Note that the inferred values of $K$ are larger when the smallest values of the two-cycle variable are discarded.  In this reduced dataset there are far fewer phase changes, which corresponds to larger values of $K$.  This consideration explains the high initial  value of $K$ in the time dependent inference shown in Fig. \ref{timeplot}a.  The initial condition in this simulation is a 50:50 mixture of high and low two-cycle values of the Ricker map.  Since there are initially no small values of the two-cycle variable, phase changes are again  unlikely and the inferred $K$ is large.  Finally, a similar consideration explains why the forecast skill score of the Ising model improves with noise (Fig.\ \ref{forecastskillplot}) since at higher noise it is more likely for a subpopulation to change phase without its amplitude of oscillation becoming small.

Figure \ref{time}b shows the observed distribution the two-cycle variable $M_{0,t}$ and reveals that this distribution is broad with values near zero quite common.   In principle, the two peaks in this distribution should be exactly symmetric ($M$ and $-M$ equally probable). The asymmetry in the figure is because the histogram is constructed from a finite time series ($6\times10^7$ steps) and because the rate of phase flips of the entire metapopulation is very low.

\section{Discussion}
\label{sec:discussion}
Previous work \citep{Noble2015} has demonstrated a correspondence between the critical point of the \textit{static}  Ising model and  the large scale  properties of stationary states of spatially coupled biological models with simple cyclic behavior.  
But for biological systems in general and ecological systems in particular, shorter term dynamical behavior is of much more interest. We have shown here that a simple \textit{dynamical}  Ising model can be successfully employed to understand and predict the dynamics of more complex, cyclic biological systems. Focusing on ecological metapopulation models, we found it necessary to include a memory term in the dynamical Ising model since subpopulations tend to maintain their phase of oscillation. 
We inferred the parameters of the dynamical Ising models that best represent the  simulated dynamics of several metapopulation models. Comparing the Ising representations to the full models, we found good agreement for both stationary states (Figure \ref{phasetwo}) and dynamical properties (Figure \ref{noneighbor}).    

A key issue in understanding ecological dynamics is prediction given the level of knowledge of a system \citep{Dietze2018}.  A simplified representation will obviously omit details but a comparison is vital for understanding the limits to prediction. We are interested in the Ising model specifically because it is a simplification of complete ecological dynamics and therefore, it is unsurprising that the forecast skill score for the Ising model is low relative to the metapopulation model (Figure \ref{forecastskillplot}). Nevertheless, the forecast skill shows that information useful for prediction is obtained, even for the simplest model.

Our results point to avenues for including more biological detail that would provide further understanding by dealing with obvious limitations  resulting from simplifications. A major insight from our work is that both static and dynamic properties of a wide range of two-cycle metapopulation models can be  reproduced by the Ising model with memory.  This means that certain features of ecological metapopulations with local 2-cycles will arise independent of the details governing the dynamics.  In other words, the success of the Ising representation reveals which aspects of ecological synchrony are detail-independent, yielding a much more general understanding of synchrony and pattern formation than can be derived from any specific model.
The dynamical Ising model with memory therefore serves as a simple baseline model with which to study ecological oscillators, without requiring any specific details of local dynamics. This is useful because we rarely know the exact structure of density dependence governing the dynamics of real populations.  Even when we do know these functions with reasonable certainty, we may lack precise parameter estimates.  By fitting the Ising model with memory to observed dynamics, we gain a quantitative representation that can be used both for understanding and prediction despite these sources of uncertainty.

Many of the more biological details we do not include here can be included in models that are only moderately more complex and still quite general. While the Ising model itself is limited to representing systems whose local dynamics can be categorized into two distinct states, such as coupled noisy 2-cycles, other spin models may allow us to take this approach further. For instance, by retaining slightly more information, the three-state Blume-Capel model shows promise for representing the general mechanism of changes in the phase of oscillation of subpopulations in a metapopulation. 
 
Heterogenous metapopulations  with varying local dynamics, noise, and dispersal on imperfect lattices could also be represented with a corresponding dynamical    Ising models with different coupling and memory at each lattice site.
The performance of the Ising representation in this more realistic setting is an important question for future research.   

Although we considered only spatially uncorrelated noise here, correlated environmental stochasticity, the Moran effect, is thought to be an important synchronizing force in ecology \citep{Moran:1953tr}.  Correlated noise can be represented by a dynamical Ising model with an external field that acts on the entire lattice.  Whether the parameters of an Ising model with such a field can be reliably estimated from simulated metapopulations with correlated noise remains an open question.  If so, Ising models fitted to data might be useful for determining the relative roles of dispersal and correlated noise in driving observations of synchrony. 

It is important to emphasize that while we have focused here on ecological models,  the insights apply much more broadly to the application of ideas from statistical physics for understanding spatially coupled biological dynamics.  A key general conclusion is that a dynamical Ising model with memory, only slightly more complex than the standard Ising model, can both represent dynamics on biologically relevant timescales and highlight the importance of  local (in space) memory of system state as a difference between biological systems and the standard Ising model. 

\section{Acknowledgements}
We thank Larry Abbott and Andrew Noble for useful discussions. The work was supported in part by NSF grant 1840221.

\renewcommand\refname{Bibliography}
\bibliographystyle{unsrt}
\bibliography{references}

\newpage
    \pagenumbering{arabic}
\setcounter{page}{1}
\renewcommand{\thefigure}{\hbAppendixPrefix\arabic{figure}}
\setcounter{figure}{0}
\renewcommand{\theequation}{\hbAppendixPrefix\arabic{equation}}
\setcounter{equation}{0}

\appendix
\begin{appendices}
\section{Metapopulation Models A - D} \label{sec:rickerNDR}
The metapopulation Models, A and B, have the Ricker map as local dynamics with the sequence $\mathbb{NDR}$.  The full dynamical equation for each subpopulation $i$ at time $t$ is, 
      \begin{equation}\label{ndr}
             X_{i,t+1} \equiv (\mathbb{NDR} \mathbf{X}_t)_i = \Big\{(1-\epsilon) X_{i,t} \exp[r(1-X_{i,t})]  + \frac{\epsilon}{4} \sum_{\langle j;i \rangle} X_{j,t} \exp[r(1-X_{j,t})]\Big\} \exp(\lambda \zeta_{i,t}).
         \end{equation}
        The growth parameter $r$ for Model A is 2.2 and for Model B is 2.4.  Both growth parameters are in the two-cycle region.
        
    Metapopulation Model C has the logistic map as local dynamics with the sequence $\mathbb{NDR}$. The full dynamical equation is, 
    \begin{equation}
             X_{i,t+1} \equiv (\mathbb{NDR} \mathbf{X}_t)_i = \Big\{(1-\epsilon) r  X_{i,t} (1-X_{i,t}) + \frac{\epsilon}{4} \sum_{\langle j;i \rangle} r X_{j,t} (1-X_{j,t})\Big\} \exp(\lambda \zeta_{i,t})
         \end{equation}
        where the growth parameter $r=3.2$.
    
    Metapopulation Model D has the Ricker map as local dynamics with growth parameter $r=2.2$ but with a different sequence, $\mathbb{NRD}$ from Modal A. The full dynamical equation is, 
      \begin{equation}
             X_{i,t+1} \equiv (\mathbb{NRD} \mathbf{X}_t)_i = \Big\{[(1-\epsilon) X_{i,t} + \frac{\epsilon}{4} \sum_{\langle j;i \rangle} X_{j,t}] \exp\big[r(1-[(1-\epsilon) X_{i,t} + \frac{\epsilon}{4} \sum_{\langle j;i \rangle} X_{j,t}])\big] \Big\} \exp(\lambda \zeta_{i,t}).
         \end{equation}

        In addition to the choice of growth parameter $r$ of the local dynamics and the sequence of processes, each metapopulation model has three other parameters: the number of patches, $N$, dispersal, $\epsilon$, and  noise, $\lambda$.
        
\begin{figure}[h]
            \centering
            \includegraphics[width=0.6\linewidth]{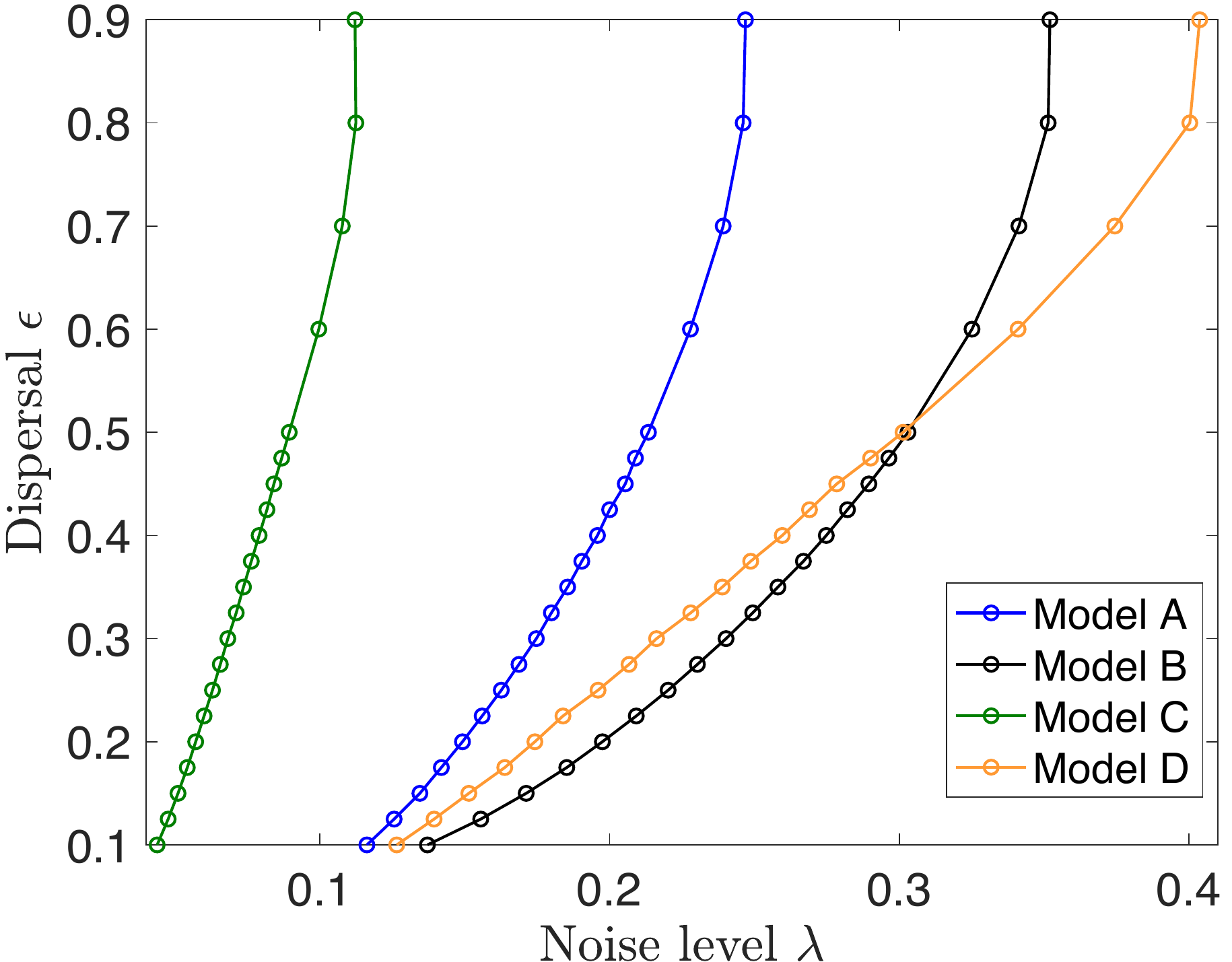}
            \caption{State diagram separating synchronous (low noise and high dispersal) and incoherent regions (high noise and low dispersal) for  the metapopulation Models A-D studied here. }
            \label{fig:phase_AD}
        \end{figure}

        The state diagram separating synchronous and incoherent regions in dispersal and noise plane ($\epsilon-\lambda$) for metapopulation Models A-D is shown in Fig. \ref{fig:phase_AD}.
        
        \section{Binder cumulant method}
\label{sec:Binder}
The state diagram for the metapopulation Models A-D is plotted by finding the critical noise $\noise$ for various dispersal values, $\epsilon$ using the Binder cumulant method. The Binder cumulant is a fourth-order cumulant of synchronization variable,
\begin{equation}
    U=1-\frac{\langle \tilde{s}_t^4 \rangle}{3 \langle \tilde{s}_t^2 \rangle ^2}
\end{equation}
where $\tilde{s}_t=\frac{1}{N} \sum_i \tilde{S}_{i,t}$ is the instantaneous synchronization variable and $\langle .\rangle$ computes the long-time average. In the synchronous region, the Binder cumulant takes the value 2/3 whereas it takes the value zero in incoherent region. At the critical transition, the critical Binder cumulant on a 2D lattice with periodic boundary conditions takes the value $U^* = 0.6169$ \citep{Selke2006}. 

\begin{figure}
    \centering
    \includegraphics[width=0.7\linewidth]{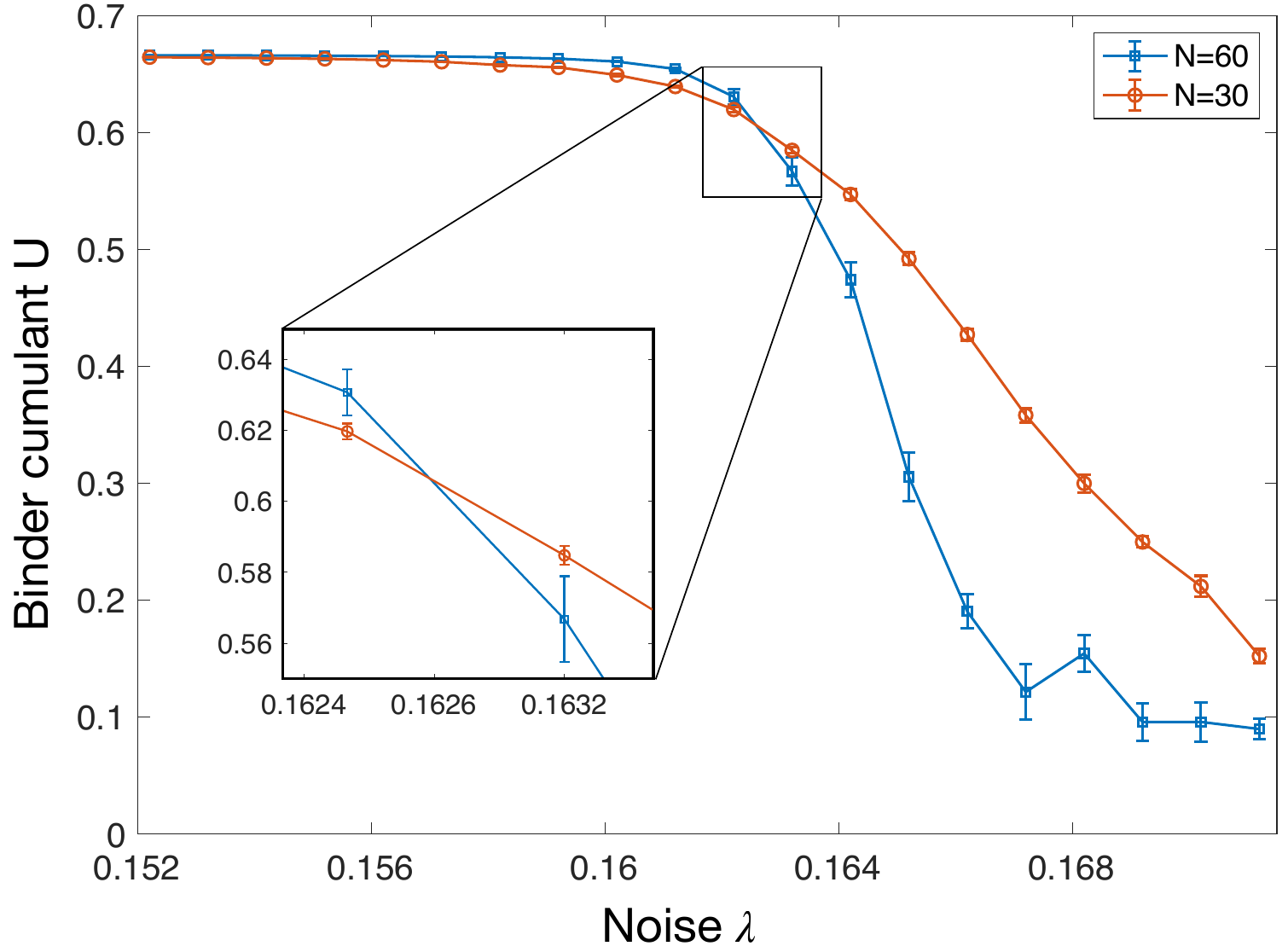}
    \caption{The critical noise $\noise$ for dispersal $\epsilon = 0.25$ is calculated for metapopulation Model A as the crossings of the Binder cumulant curves for patch sizes $N=30,60$. The inset shows the crossings of the Binder cumulant curves at $\noise = 0.1626$. 
    }
    \label{fig:Binder}
\end{figure}

The critical noise $\noise$ for a given dispersal $\epsilon$ is obtained from the crossings of the Binder cumulant curves plotted as function of noise for different number of patches, $N$. In this work, we use $N=30,60$ to find the critical noise from the crossings of the Binder cumulant curve. The Binder cumulant curves are shown for dispersal $\epsilon=0.25$ as function of noise $\lambda$ in Fig. \ref{fig:Binder}. The Binder cumulant curves crosses at critical noise $\noise= 0.1626$ as shown in the inset. 

\section{Dynamical Ising model with memory}
\label{app:ising}

The dynamical Ising model with memory \citep{CIRILLO201436, Pre} has nearest neighbor coupling $J$ and a self-interaction strength $K$. The single spin transition probability,  $P(S_{i,t+1}|\mathbf{S}_{t},J,K)$, is given by,
\begin{equation}\label{probmemory}
    P(S_{i,t+1}| \mathbf{S}_{t},J,K) =  \frac{\exp\big[(J h_{i,t}+ K S_{i,t}) S_{i,t+1}\big]}{2\cosh\big[J h_{i,t}+K S_{i,t}\big]} 
\end{equation}
where the local field, $h_{i,t}$, is the sum of the nearest neighbors of $S_{i,t}$ at time $t$
\begin{equation}
    h_{i,t} = \sum_{\langle j;i \rangle} S_{j,t}.
\end{equation}

The transition probability for the full spin configuration for parallel dynamics (in which all spins are updated simultaneously each time step) is given by,
\begin{equation}\label{parprob}
    P(\mathbf{S}_{t+1}|\mathbf{S}_{t},J,K)= \prod_i P(S_{i,t+1} | \mathbf{S}_{t},J,K)=\prod_i \frac{\exp[(J h_{i,t}+K S_{i,t}) S_{i,t+1}]}{2\cosh[J h_{i,t}+K S_{i,t}]}.
\end{equation}
Note that the memory ($K$) and nearest neighbor coupling ($J$) parameters are both dimensionless. It is also common to write these parameters as ratios, $J/T$ and $K/T$ where now $J$ and $K$ are considered interaction ``energies'' and $T$ is a ``temperature" and reflects the noise level in the system. Because only the ratios appear in Eq.\ \eqref{parprob}, it is not possible to distinguish the difference between increasing noise and decreasing interaction strength, without making additional assumptions. 

The transition probability for the standard dynamical Ising model is obtained by setting $K=0$ so the transition probability for  spin, $S_{i,t+1}$ depends only on its neighbors  and not on itself. This is in contrast to ecological dynamics, in which the current local state is a density dependent function of the previous one.

\begin{figure}
    \centering
    \includegraphics[width = .5\textwidth]{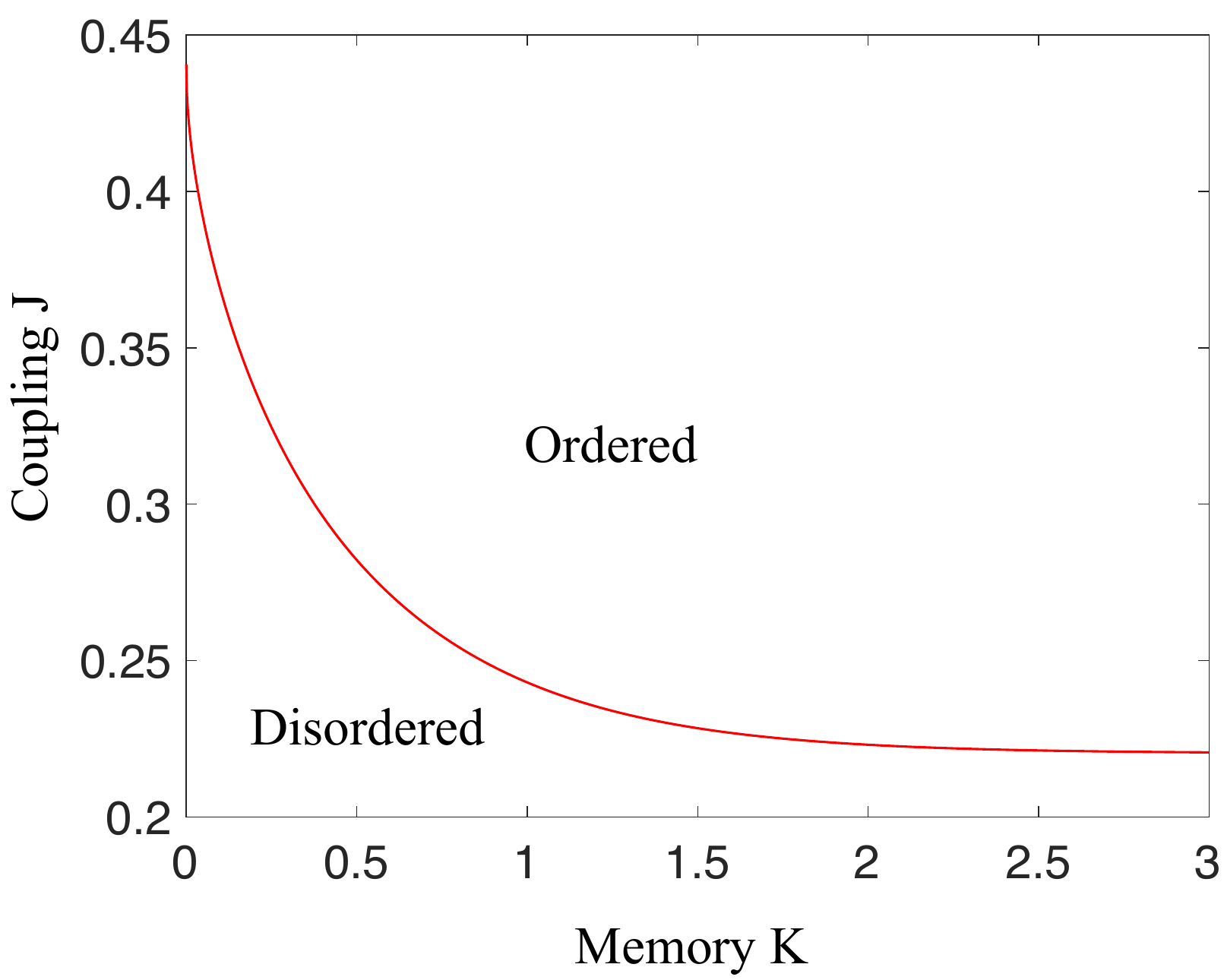}
    \caption{State diagram of the dynamical Ising model with memory $K$ and nearest neighbor coupling $J$ \citep{Pre}.   The solid red line is the critical line separating the ordered (synchronous) and disordered (incoherent) regions in the $J$--$K$ plane. The tendency toward ordering increases with both $J$ and $K$. The point $K=0$, $J_c(0)=0.44\ldots$ is the critical point of the standard Ising model. At large $K$, the critical coupling approaches half the value of the standard Ising model. }
    \label{Ising phase}
\end{figure}

The Ising model with memory undergoes a critical transition in the Ising universality class from a disordered to an ordered state.  For a fixed value of $K$ this transition occurs at a value, $J_c(K)$. The critical line, $J_c(K)$, separating ordered and disordered states in the coupling-memory ($J$--$K$) plane is shown in Fig.\ \ref{Ising phase} \citep{Pre}.

\section{Inference methods}
\label{sec:inference}
In this section we introduce maximum likelihood inference methods to determine the parameter values $\{J,K\}$ for the dynamical Ising model with memory that best describe the dynamics of the metapopulation model. Inferring an Ising model from data is called an inverse Ising problem \citep{Aurell:2012}. Inverse Ising problems typically refer to inferring parameters of the stationary Gibbs distribution that are most like the data sampled at a single time. By contrast, here we use {\it dynamical} inference \citep{Decelle:2016} to find the dynamical Ising model that best represents the dynamics of the metapopulation model. Of course, if the dynamical model is accurate, its stationary states should be close to the stationary states of the metapopulation.

The likelihood that we maximize is related via Bayes' theorem and several assumptions to the transition probability of the Ising model, $P(\mathbf{S}_{t+1}|\mathbf{S}_t,J,K)$ 
(see Eq. \eqref{parprob}). 

We are interested in the likelihood $L\big(\{J,K\}|\mathbf{\tilde{S}}_{t+1},\mathbf{\tilde{S}}_{t}\big)$ of the parameter values \{J,K\} given spin configurations from two successive time steps, $\{\mathbf{\tilde{S}}_{t+1},\mathbf{\tilde{S}}_{t}\}$ of the metapopulation.  Using Bayes' theorem and the assumption of uniform priors for both the parameters and the spin configurations, the likelihood is proportional to the transition probability of the Ising model with memory, $P(\mathbf{S}_{t+1}|\mathbf{S}_{t},J,K)$ (Eq.\ \ref{parprob}). 
It is equivalent but more convenient to maximize the logarithm of the likelihood, $\mathcal{L}= \log L$,
\begin{equation}\label{loglike}
    \mathcal{L}\big(\{J,K\}|\mathbf{\tilde{S}}_{t+1},\mathbf{\tilde{S}}_{t}\big) \propto \sum_i \Big\{[J \tilde{h}_{i,t}+K \tilde{S}_{i,t}] \tilde{S}_{i,t+1}-\log [2\cosh(J \tilde{h}_{i,t}{S}_{i,t}+K \tilde)]\Big\}.
\end{equation}
Finally, while this is an exact formula for inferring parameters from data, we assume that maximizing $\mathcal{L}$ using the Ising representation of the metapopulation data $\mathbf{\tilde{S}}_{t}$ will yield a good Ising model representation.
Thus, the goal is to maximize 
$\mathcal{L}\big(\{J,K\}|\mathbf{\tilde{S}}_{t+1},\mathbf{\tilde{S}}_{t}\big)$.  Because the Ising spin representation of the metapopulation involves only binary variables and because of additional symmetries of the dynamical Ising model,  maximizing the likelihood reduces to a simple numerical fit of the two parameters with respect to ten numbers obtained directly from the data.

Possible values of the product $h_{i,t}S_{i,t}$ for all different spin initial values at time $t$ are $\{-4,-2,0,2,4\}$ and for $S_{i,t}S_{i,t+1}$ are $\{-1,1\}$. Considering $\mathcal{Z}_2$ symmetry of the model, there are only 10 different possible values for each term in the sum in  Eq.\ \eqref{loglike}. Also, the sum of probabilities (Eq.\ \ref{parprob}) with the two possible values of $S_{i,t}S_{i,t+1}$ with same $h_{i,t}S_{i,t}$ must be one. 
This can be understood as given the product $h_{i,t}S_{i,t}$, spin $S_{i,t+1}$ either flips or not which is equivalent as $S_{i,t}S_{i,t+1}$ being either $-1$ or $1$. 

The flip probability given the product is $P_f(h_{i,t}S_{i,t})$ as shown in Eq.\ \eqref{eqn:flip}. Whenever $S_{i,t}S_{i,t+1} = -1$, the corresponding probability is $P_f(h_{i,t}S_{i,t})$ and when $S_{i,t}S_{i,t+1} = 1$ the probability is $1-P_f(h_{i,t}S_{i,t})$.

Now, spins in $\mathbf{\tilde{S}}_{t}$ can be grouped into 5 bins corresponding to their products $h_{i,t}S_{i,t}$. Lets say the number of spins in each bin is $n_k$ where $k$ represent the values of the product $h_{i,t}S_{i,t}$ ($\{-4,-2,0,2,4,\}$) then $\sum _k n_k = N$ with $N$ being the total number of spins. Out of these $n_{k}$ spins in a bin, $n_{k}^{f}$ spins chose to flip at time $t+1$ with $S_{i,t}S_{i,t+1}=-1$ which enforces $n_{k}-n_{k}^{f}$ spins with no flip at $t+1$. 
Then the above log-likelihood function in Eq. \eqref{loglike} can be rewritten as,
\begin{equation}
    \mathcal{L}\big(\{J,K\}|\mathbf{\tilde{S}}_{t+1},\mathbf{\tilde{S}}_{t}\big) =\sum_{k} \Big\{n_{k}^{f} \log(P_f(k)) + [n_{k}-n_{k}^{f}] \log(1-P_f(k))\Big\}
\end{equation}
Here $\{P_f(k)\}$ are functions of parameters $\{J,K\}$ and $n_{k}$s are specific to the spin configurations $\mathbf{\tilde{S}}_{t},\mathbf{\tilde{S}}_{t+1}$. So given the spin configurations at time $t$ and $t+1$, the log-likelihood is a function of $\{J,K\}$ which is maximized to infer parameters.

The spin configurations $\mathbf{\tilde{S}}_{t},\mathbf{\tilde{S}}_{t+1}$  are obtained over 100 independent runs and the $n_{k}$s  are collected which form a multinomial distribution.  The error bars for the observations are calculated by the bootstrap method \citep{newmanb1999} on the obtained multinomial distribution.  The error bars on the inferred results in Fig. \ref{phasetwo} contain only statistical errors. The systematic errors in identifying exact parameters of the metapopulation critical line are not included.

\section{Forecast skill}
\label{sec:forecastskill}
Forecast skill measures the accuracy of a predictive model. We use the Brier skill score to characterize the ability of the inferred dynamical Ising model to predict the metapopulation model one step in the future. 
The Brier score, $BS$ is the mean squared error in a probabilistic prediction, $P_i$ of a binary observable, $F_i$ averaged over many events $k$.  
\begin{equation}
    BS =  \frac{1}{M} \sum_{k=1}^M (P_k - F_k)^2.
\end{equation}
Here $F_k$ is either zero or one and $P_k$ the predicted probability that $F_k$ will be one.
For example, in weather forecasting, a binary observable is rain ($F_k=1$) or no rain ($F_k=0$) and $P_k$ the probabilistic forecast (e.g.\ 20\% chance of rain).

Brier forecast skill, $FS$ compares the Brier score of the probabilistic prediction to a simple reference forecast,  
\begin{equation}\label{eq:forecastskill}
    FS = 1- \frac{BS}{BS_{\text{ref}}}
\end{equation}
where $BS_{\text{ref}}$ is the Brier score of the reference forecast (e.g.\ the overall climatological probability of rain).  Forecast skill is 1 for a perfect forecast, greater than zero for a forecast that improves on the reference, and negative for forecast with less skill than the reference.

We seek to  predict the probability that a subpopulation changes its phase of oscillation (``flips'') in the next time step.
Given successive two-cycle configurations for a given dispersal and noise, $\mathbf{M}_t$ and $\mathbf{M}_{t+1}$, we obtain the corresponding phase configurations, $\mathbf{\tilde{S}}_t$ and $\mathbf{\tilde{S}}_{t+1}$ (see Eq.\ \eqref{mapping}).  A phase change of subpopulation $i$ occurs at time $t$ if the phase flip variable, $F_{i,t}$,  
\begin{equation}
    F_{i,t} = \frac{|\tilde{S}_{i,t+1}-\tilde{S}_{i,t}|}{2},
\end{equation}
is one and no phase change occurs if it is zero.

The probabilistic prediction, $P_{f}$ is made using the flip probability of the inferred Ising model with memory,  $P_{f}=P(S_{i,t+1}=-S_{i,t}| \mathbf{S}_{t},J,K)$ (see Eq. \eqref{eqn:flip}) where the parameters {$J,K$} are taken from the inference results for the chosen dispersal and noise and $\mathbf{\tilde{S}}_{t}$ is set by the metapopulation phase variables $\mathbf{S}_{t}$.   
The Brier score for the dynamical Ising model with memory is obtained by averaging over the $N=60^2$ subpopulations of the metapopulation and then over 100 independent samples of the metapopulation in the stationary state. For each subpopulation, the product $\tilde{h}_{i,t}S_{i,t}$ is the input to flip probability.  The flip probabilities are shown in Fig.\ \ref{noneighbor}, panels (b)-(f).   The reference prediction is the overall probability of a flip averaged over the stationary state.  This probability is plotted in Fig.\ \ref{noneighbor}, panel (a).

Since the metapopulation models are stochastic, the skill scores of the models themselves are less than one.  In order to calculate the Brier score of the metapopulation models we use the defining lattice maps, for example Eq.\ \eqref{ndr} for Models A and B.  For given configurations $\mathbf{X}_t$ and $\mathbf{X}_{t+1}$ we can solve an algebraic equation for a bound on the noise factor $\exp(\lambda \zeta_{i,t})$ for which a flip occurs for a given subpopulation. The probability that this bound is satisfied can be calculated analytically as an error function and is the prediction of the model. The forecast skill score of the model itself is shown as the upper curve in Fig.\ \ref{forecastskillplot} and is a bound on the performance of any other predictive model of the metapopulation.

\section{Simulation details}
\label{sec:simdetails}

We studied the dynamics of the four metapopulation models (Table \ref{tab:four_models}) numerically on a square lattice of length 60 ($N=60^2$) with periodic boundary conditions. 
For a given metapopulation model, we have carried out simulations for various values of dispersal $(0<\epsilon < 1)$ with noise $\noise$  chosen such that the parameters are on the critical line (see Fig. \ref{phase plot} for metapopulation model A). 
Three consecutive lattice configurations were collected for analysis at the end of each run which consists of  $3 \times 10^6 $ timesteps.
Results shown in the following sections were obtained from the average over 100 individual runs, corresponding to 100 independent sets of three consecutive metapopulation configurations. 

The best fit parameters values $\{J,K\}$ of the dynamical Ising model with memory were obtained using maximum likelihood inference methods (Appendix \ref{sec:inference}). The likelihood is proportional to the transition probability of the Ising model with memory, $P(\mathbf{S}_{t+1}|\mathbf{S}_{t},J,K)$ (see Eq.\ \eqref{parprob} of Appendix \ref{app:ising}).

\end{appendices}

\end{document}